\documentclass[a4paper,11pt]{article}
\usepackage{jinstpub} % for details on the use of the package, please see the JINST-author-manual
\usepackage{lineno}
\usepackage{siunitx}
\usepackage{hyperref}
\usepackage{subcaption}%, subcaption
\usepackage{multicol}

\DeclareSIUnit\barn{b}

\title{\boldmath Design of a thermal loading resilient optical enhancement cavity for operation at 515~nm for X-ray production through inverse Compton scattering in an energy recovery linac}

\author[a]{Alice Renaux}
\author[a]{Aurélien Martens}
\author[a]{Yann Peinaud}
\author[a]{Ronic Chiche}
\author[a]{Kevin Dupraz}
\author[a]{Marie Jacquet}
\author[a]{Daniele Nutarelli}
\author[a]{Fabian Zomer}
\affiliation[a]{Laboratoire de Physique des 2 Infinis Irène Joliot-Curie (IJCLab), Université Paris-Saclay, CNRS/IN2P3\\ Orsay, France}

\emailAdd{alice.renaux@ijclab.in2p3.fr}

\abstract{A fast and simple method to optimize a high-power optical enhancement cavity is proposed. It is applied to a four-mirror bow-tie cavity operating at 515~nm, which is to be implemented within the PERLE ERL for the production of X-rays through inverse Compton scattering. The optimized figure of merit is the expected X-ray photon rate. Thermal loading is carefully evaluated, giving also a path towards very high average power operation in infrared. One shows that photon rates exceeding $10^{11}$ per second at average energies above 280~keV in apertures of a few miliradians are within reach. For easier comparison with existing Compact X-ray sources providing high rates of photons, the performance of the proposed design is also evaluated for a system delivering infrared laser pulses. It is shown that a source delivering more than $10^{13}$ photons per second is achievable with this design, an unprecedented performance for existing Compton X-ray sources. The design procedure employed here can be easily applied to other applications of high-power high-finesse optical cavities. %It motivates an experimental demonstration of an optical cavity operating at hundreds of kilowatts of average power at 515~nm. For easier comparison with existing Compact X-ray sources providing high rates of photons, the performance of the proposed design is also evaluated for a system delivering infrared laser pulses. It is shown that a source delivering more than $10^{13}$ photons per second is achievable with this design, an unprecedented performance for existing Compton X-ray sources. The procedure employed here can be easily applied to other applications of high-power high-finesse optical cavities.
}

\begin{document}
\maketitle
\flushbottom

\section{Introduction}
\label{seq:introduction}

Various fields of fundamental and applied research utilize high-power optical enhancement cavities (OECs). Fundamental physics applications of OECs include the detection of gravitational waves in large-scale Michelson interferometers \cite{abbott2016gw150914,acernese2014advanced,akutsu_kagra_2019, Willke_2006} and searches for axion-like particles \cite{B_hre_2013, PhysRevLett.132.191002}.%, and efforts to measure magnetic birefringence \cite{EJLLI20201}. 
The scope of the Gamma Factory project at CERN covers fundamental physics applications from atomic physics to particle physics, accelerator physics, and applied nuclear physics \cite{krasny2015gammafactoryproposalcern, PhysRevAccelBeams.25.101601}. Prospects for applied physics applications of OECs include nuclear fusion research, such as the photo-neutralization of negative ion beams \cite{simonin2016negative} or laser-based inertial fusion systems \cite{sunahara2025laser}. %OECs are currently in use in non-linear optics, for the exploitation of the optical rectification process \cite{Canella:25} or of the high harmonic generation process (HHG) for X-ray or EUV-ray beams production or for attosecond pulses production \cite{pupeza2021extreme}. 
The development of high brilliance and high repetition rate EUV-ray sources for lithography is also being investigated, based on the steady-state microbunching of electron beams by OEC-enhanced lasers \cite{deng2021experimental, kruschinski_confirming_2024}. Inverse Compton scattering sources (ICS) enable the production of highly monochromatic, energy-tunable X-ray or $\gamma$-ray beams \cite{eggl2016munich, DUPRAZ2020100051}, where a high-energy accelerator is coupled to a laser system to provide high-energy photons for diverse applications.

%Several ICS projects employ room-temperature linac and powerful laser systems with modest photon rates \cite{graves:linac2024-weya004,VanElk:25, SAMSAM2024168990,AMOUDRY2025170287,10.3389/fphy.2024.1472759}. Higher yields of photons are achieved with electron rings coupled to OECs \cite{Gunther:ok5018,jacquet2024first} where a very high average power, close to one megawatt can be stacked \cite{carstens2014megawatt, lu2024stable,700kW}. Developments of burst mode laser systems, either exploiting optical recirculators \cite{CHALEIL2016113,PhysRevSTAB.17.033501} or OECs operated in burst regime, have emerged over the past decade \cite{sakaue_stabilization_2018,PhysRevAccelBeams.21.121601}. Another compromise lies in exploiting OECs with cryogenic electron accelerators, as Energy Recovery Linacs (ERLs). It has been realized at the cERL of KEK \cite{nagai2015demonstration} and is considered for the PERLE project at IJCLab \cite{Angal-Kalinin_2018,martens2021towards}.

It is a striking fact that ICS linac projects often employ green laser systems to optimize the source footprint and cost, as the energy of the scattered photons scales quadratically with the electron energy and linearly with the initial photon energy. However, high-power OECs, with more than 100~kW, were never operated around 515-530~nm. To the best of our knowledge, about 4~kW was reached in a two-mirror spherical cavity seeded by a continuous-wave (CW) green laser for Compton polarimetry \cite{RAKHMAN201682}. It is two orders of magnitude less than what was achieved in infrared~\cite{carstens2014megawatt, lu2024stable, 700kW}. Such a power reduction spoils the advantage of high repetition rates of an electron storage ring. Improving the performance of high-power OECs operated in green for ICS is of interest, for instance, for an energy upgrade of MuCLS \cite{10.1117/12.3042333}, or to upgrade existing ICS linacs with burst mode OECs \cite{graves:linac2024-weya004,VanElk:25, SAMSAM2024168990,AMOUDRY2025170287,10.3389/fphy.2024.1472759}.

Seed laser systems delivering a few hundred Watts are available in green~\cite{Zhao:17}. As a consequence, this paper is focused uniquely on the OEC design for a high finesse high power at 515~nm, never been reported in the literature, and not on the seed laser design. Thermal loading in the OEC is a known subject~\cite{carstens2014megawatt,Bullington:08}, however, to the best of our knowledge, no detailed study focused on the optimization of bow-tie OEC performance under thermal load for high-finesse, high-power operation at \qty{515}{\nm} is available in the literature.
We show, in this paper, that high performance can be obtained at high power by (i) optimizing the geometry ignoring thermal effects; (ii) choosing the radius of curvature of the spherical mirrors by minimizing fundamental mode deformation due to thermal expansion; (iii) recovering coupling by proper design of a simplified matching telescope at high-average power. We examplify it by applying this to the design of a system for the PERLE project. However, this simple procedure remains valid for other applications where the implementation of high-power OEC is required.
%we do not aim at discussing here the design and delivery of a high-power laser amplifier operated in green.

%In the past decades, difficulties related to the high thermal load in high-finesse high-average-power OECs have been overcome. Design constraints due to thermal loading in optical cavities were first studied \cite{carstens2014megawatt}. Later on, higher-order mode dampers have been introduced \cite{Amoudry:20} to mitigate mode-degeneracies \cite{Bullington:08}. It has been more recently accounted for by design in the Gamma Factory proof-of-principle experiment \cite{PhysRevAccelBeams.25.101601}. More recently, it has been realized that thermal lensing may also significantly impact the ability to achieve very high average power in OECs \cite{700kW}. Contrary to OECs optimized for interaction with ion beams \cite{PhysRevAccelBeams.25.101601}, ICS requires a focused laser beam for optimized interaction with electron beams \cite{DUPRAZ2020100051}. As a consequence, a four-mirror bow-tie geometry is generally implemented, as it provides independent tuning of the OEC round-trip length and laser focusing \cite{Zomer:09}. This property also allows for a compromise in the laser beam size, resulting in lower sensitivity to thermal effects \cite{carstens2014megawatt,lu2024stable}.

This paper is thus devoted to the design of a four-mirror bow-tie cavity intended for operation at high average power for the PERLE project. It is designed to sustain tight accelerator integration constraints and thermal effects by choice of the optical cavity parameters. The design method is described in \autoref{sec:design}. A particular emphasis is placed on designing a system that is robust against the detrimental effects of thermal loading, thereby preserving OEC enhancement capabilities. The performance of the ICS is then anticipated in \autoref{sec:perf}. It is provided for both near-infrared and green wavelengths.

\section{Design of an OEC for the PERLE ERL}\label{sec:design}
Besides technical aspects of the integration of an OEC in an accelerator~\cite{Bonis_2012,DUPRAZ2020100051}, it is necessary to design an OEC that can be integrated within the accelerator and be tolerant to thermal load in the mirrors, induced by the necessary high average power. This section is devoted to the description of the method we have developed to reach both of these goals. This is done in two steps. First, we intend to optimize the cavity geometrical parameters accounting for integration constraints. Second, we choose the radius of curvature of the spherical mirrors to minimize the sensitivity of the cavity to the thermal load and preserve performance despite the high average power. The figure of merit that we choose to optimize is the production rate of X-ray photons $\frac{dn_{\text{X}}}{dt}$. It reads 
\begin{equation}
    \label{eq:flux}
	\frac{dn_{\text{X}}}{dt}=\sigma_{\text{C}}\times\Upsilon\times\frac{Q}{e}\times\frac{P_{\text{cav}}}{h\nu_{\text{0}}},
\end{equation}
where $\sigma_{\text{C}}\simeq\qty{0.67}{\barn}$ is Compton cross-section, $Q$ the charge of an electron bunch, $e$ the elementary charge, $h\nu_{\text{0}}$ the energy of one incident photon, $P_{\text{cav}}=U\times f_{\text{rep}}$ the average intracavity power with $U$ the energy of a laser pulse and $f_{\text{rep}}$ the laser repetition rate, that we chose equal to the electron beam repetition rate \cite{suzuki1976general}. The term $\Upsilon$ denotes the spatio-temporal electron bunch-laser pulse overlap and reads 
\begin{equation}
    \label{eq:upsilon}
	\Upsilon=\frac{1}{\sqrt{\sigma_{\text{l, y}}^{2}+\sigma_{\text{e, y}}^{2}}\sqrt{(\sigma_{\text{l, x}}^{2}+\sigma_{\text{e, x}}^{2})+(\sigma_{\text{l, z}}^{2}+\sigma_{\text{e, z}}^{2})\tan^{2}{\frac{\theta_{\text{c}}}{2}}}},
\end{equation}
where it is assumed that the electron bunch and laser intensity profiles follow a normal distribution and $\sigma_{\text{a, b}}$ denotes the electron ($a=e$) and laser intensity ($b=l$) RMS size in the horizontal ($b=x$), vertical ($b=y$) and longitudinal ($b=z$) directions. Both beams are assumed not to be focused too strongly to neglect the luminosity reduction due to the hourglass effect. For the purpose of the design, this effect is ignored and validated after the laser beam parameters are known. The crossing angle between the electron and laser beams is denoted $\pi-\theta_{\text{c}}$. The beams are nearly contra-propagative as $\theta_{\text{c}}\ll \pi$. By convention, we assume that the PERLE ERL lies in the horizontal plane $(x,z)$.

\subsection{Geometry optimization}
\begin{figure}[htbp]
    \centering
    \includegraphics[width=0.75\textwidth]{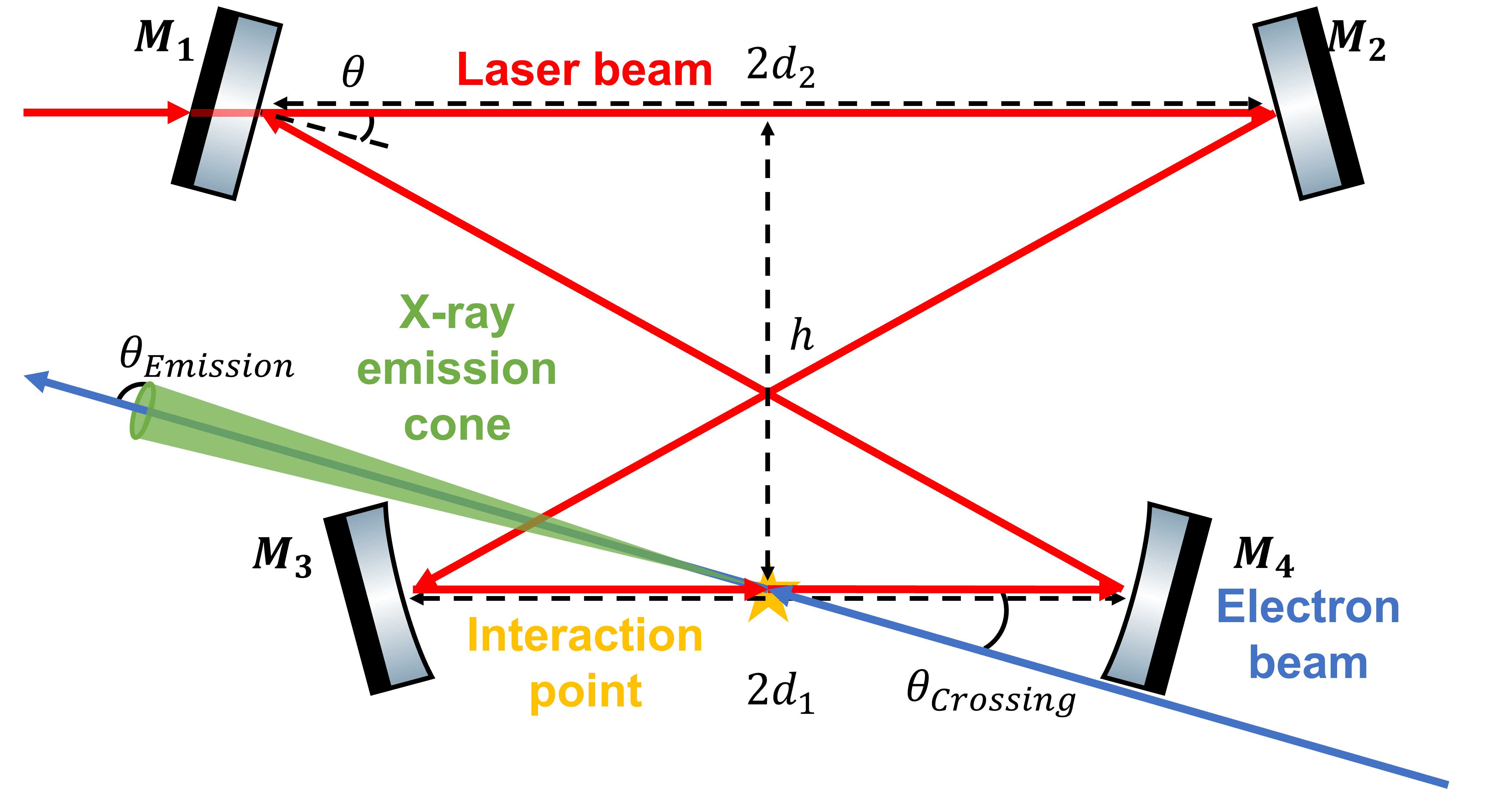}
    \caption{Parametrization of a two-mirror bow-tie OEC. $M_{\text{1}}$ is the planar input mirror, $M_{\text{2}}$ is the planar end-of cavity mirror and $M_{\text{3}}$ and $M_{\text{4}}$ are the spherical end-of-cavity mirrors. The interaction point with the electron beam is located in the middle of the $M_{\text{3}}-M_{\text{4}}$ segment.}
    \label{fig:cavity_parameters}
\end{figure}

We assume in the following that the OEC is planar and lies in the same plane, which significantly facilitates its integration. ICS requires a focused laser beam for optimized interaction with electron beams \cite{DUPRAZ2020100051}. As a consequence, a four-mirror bow-tie geometry is chosen, as it provides independent tuning of the OEC round-trip length and laser focusing \cite{Zomer:09}. The OEC geometry is depicted in \autoref{fig:cavity_parameters}, where the path of the laser inside the OEC is shown along the direction of the electron beam and the $1/\gamma$ cone of emission of scattered photons. We impose a symmetry axis for the geometry that goes through the middle points between the two planar mirrors M$_1$ and M$_2$, and the two spherical mirrors M$_3$ and M$_4$, where the electron beam interacts. We assume in the following that the mirror diameter $\phi$ is one inch. In order to optimize the geometry of the OEC, we assume the parameters for the ERL that are listed in \autoref{tab:specs_perle}. In particular, the electron beam emittance is assumed to be the same in both horizontal and vertical planes.

\begin{table}[htbp]
    \centering
    \caption{PERLE electron beam specifications.\label{tab:specs_perle}}
    \smallskip
    \begin{tabular}{l|r}
        \hline
        Repetition rate $f_{\text{rep}}$ [MHz] & \qty 40.077 \\
        Electron bunch charge $Q$ [pC] & 500 \\
        Electron beam energy $E_{\text{e}}$ [MeV] & 89 \\
        Electron beam relative energy spread $\Delta E_{\text{e}}$ [\%] & 0.1 \\
        Normalized emittance $\epsilon_{\text{n}}$ [mm$\cdot$mrad] & 3 \\
        RMS transverse beam size $\sigma_{x, e}\times\sigma_{y, e}$ [µm$\times$µm] & 75$\times$75 \\
        RMS longitudinal beam size (bunch length) $\sigma_{z, e}$ [ps] & 5 \\
        \hline
    \end{tabular}
\end{table}

\begin{figure}[htbp]
    \centering
    \includegraphics[width=0.49\textwidth]{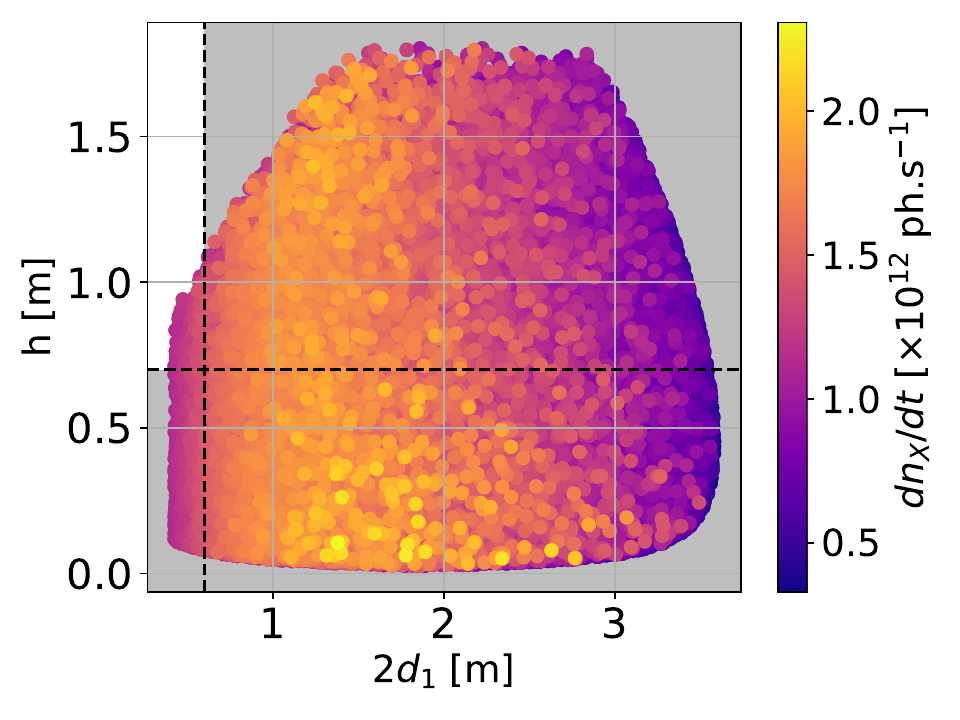}
    \includegraphics[width=0.49\textwidth]{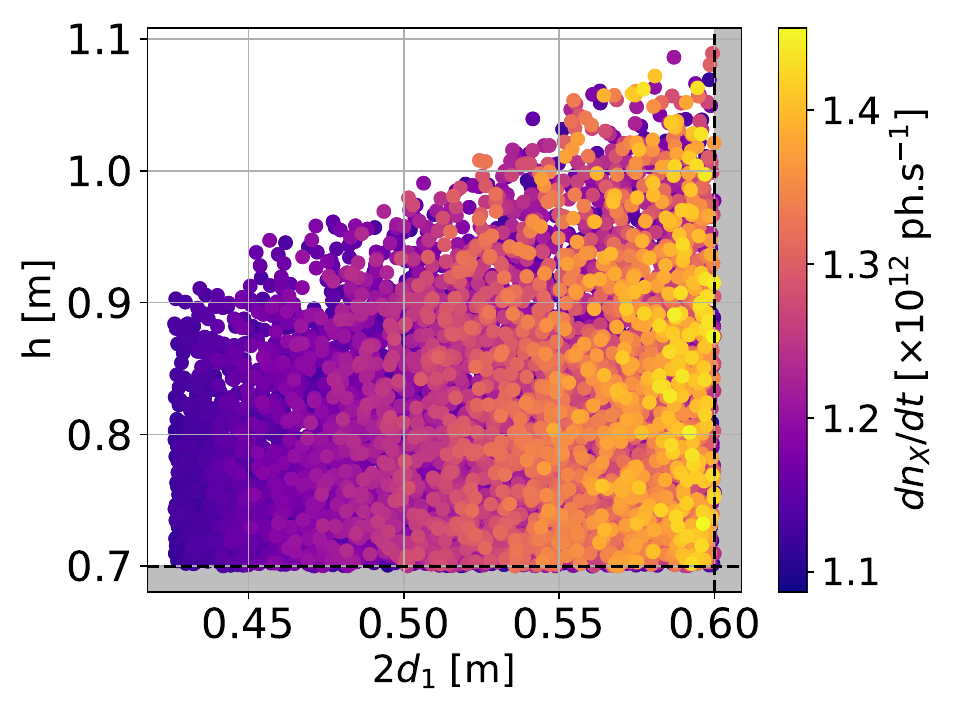}
    \caption{(Left) Expected rate of produced photons as a function of the $h$ and $2d_1$ parameters. Black dashed lines materialize integration constraints, where the forbidden zone for integration correspond the the greyed area. The allowed range is a narrow left upper corner, where the maximum achievable rate is reduced by about 40~\% with respect to the maximum value in the whole range. (Right) A zoom on the region acceptable for integration, the best performance is obtained in the right-hand side, where $2d_1\approx 0.6$~m.}
    \label{fig:Figure2_filtered}
\end{figure}

The Monte Carlo procedure that we previously employed \cite{PhysRevAccelBeams.21.121601} is followed, except that here we fix the pulse energy, as the spot size on the mirrors is relatively large and the mirror damage threshold is not a limiting factor. To start with, we decide to fix the laser pulse duration to $\sigma_{l, z}=\sigma_{e, z}$. The round-trip length of the OEC is determined by the frequency that has to remain locked to that of the ERL, $L_{\text{cav}}=\frac{c}{f_{\text{rep}}}$. The parameters $d_{\text{2}}$ and $h$ are thrown randomly with flat distributions in the ranges $[0, \frac{L_{\text{cav}}}{4}]$ for both. One then computes $d_{\text{1}}=\left(L_{\text{cav}}^2-L_{\text{cav}}d_2-4h^2\right)/(4L_{\text{cav}})$ and the incidence angle on the mirrors $\theta=\arctan{\left({h}/{(d_1+d_2)}\right)}/2$. The radius of curvature of the two spherical mirrors is then thrown in a range $[1.8,2.0]\times d_1$, where the maximum of $\Upsilon$ is found to be located. A simulation in a larger range is possible, but would only reduce the useful statistics. A tight integration constraint from the accelerator point of view can be accounted for already at this stage. In order to leave space for the electron vacuum tube to shield the mirror and its mounts from the high current of the electron beam, a minimal distance of $d_{\text{min}}$~=~37~mm is set in between the center of the spherical mirrors and the electron beam. It is setting a $d_1$ dependent largest crossing angle (minimum $\theta_{\text{C}}$) in between the two beams. As a consequence, we decide to throw $\theta_{\text{C}}$ in the range $[\arcsin{\left({d_{\text{min}}}/{d_1}\right)}, 0.17~\text{rad}]$, where the upper limit is arbitrarily chosen but felt reasonable from the integration and performance point of view. The ABCD matrix of each optical element of the OEC is then computed under the paraxial approximation of Gaussian beams \cite{kogelnik1966laser}. The size of the laser beam can then be estimated with the round-trip ABCD matrix \cite{kogelnik1966laser} at the interaction point to compute $\Upsilon$ and the photon rate. We also explicitly check that the fluence $\mathcal{F}={U}/\left({\pi w_{x}w_y}\right)$ on each mirror does not exceed a damage threshold $\mathcal{F}_{\text{max}}=\qty{1}{\J.\cm^{-2}}$, where $w_{x,y}$ represent the beam radius on the mirrors in the horizontal and vertical directions. It is also checked that the mirror spot size is sufficiently small with respect to the mirror diameter $\phi$ by requiring that $\alpha\phi>4\max{(w_x,w_y)}$, where $\alpha$ is the mirror coating clear aperture, which we set to 0.8. The result of this Monte Carlo is shown in \autoref{fig:Figure2_filtered}, where the rate of expected produced photons is shown in color for the set of $(2d_1,h)$ values generated. We see that a maximum lies in the region $2d_1\approx 1~m$. However, the space for integrating the OEC into the accelerator is very limited. A maximum distance in between the spherical mirrors of $2d_1\leq 0.6$~m is set to accommodate the presence of nearby quadrupoles. Moreover, to leave space for these and a reasonable size for the mirror's vacuum chambers imposes a minimal distance between the axes of spherical and planar mirrors such that $h\geq 0.7$~m. It affects the achievable rate of scattered photons by reducing it by about 40\%. These constraints are materialized by black lines in this figure. A second Monte Carlo simulation is conducted in this very limited phase space, yielding a result that indicates the ideal solution lies in the bottom right corner. This geometry optimization is obtained assuming that there is no thermal loading in the mirror. Anticipating the results of the next section, we have explicitly checked that including thermal loading does not change this result significantly: the geometry remains optimum in the same region of the phase space. It only affects the choice of the radius of curvature that we motivate in the next section. This can be explained by the fact that the transverse beam size of the electron beam remains large, and that the mode of the optical cavity is not significantly affected at the interaction point. This might be different in situations where the electron is more focused; in that case, the complete Monte Carlo procedure would apply. As a consequence, the cavity geometry is fixed to the parameters listed in \autoref{tab:specs_oec}. Here, the crossing angle is set at its minimum acceptable value, as determined by the previous criteria. 

\begin{table}[htbp]
    \centering
    \caption{Geometrical parameters of the OEC accounting for integration constraints. \label{tab:specs_oec}}
    \smallskip
    \resizebox{\textwidth}{!}{
    \begin{tabular}{c|c|c|c|c|c|c|c}
        \hline
        $f_{rep}$ [\unit{\MHz}] & $L_{\text{cav}}$ [\unit{\m}] & $2d_{\text{1}}$ [\unit{\m}] & $2d_{\text{2}}$ [\unit{\m}] & $h$ [\unit{\m}] & $\theta$ [\unit{\degree}] & $\theta_{\text{Crossing}}$ [\unit{\degree}] & $\phi$ [\unit{\cm}]\\
        \qty{40.077}{} & \qty{7.48}{} & \qty{0.6}{} & \qty{3.0092}{} & \qty{0.7}{} & \qty{10.6}{} & \qty{7}{} & \qty{2.54}{}\\
        \hline
    \end{tabular}}
\end{table}

\subsection{Optimization of the optical cavity behavior under thermal load}
Now that the cavity geometry is fixed, one investigates the behavior of the optical system in the presence of the unavoidable thermal load induced by residual absorption in the mirrors' coatings. Thermal effects in OECs have been identified for a long time \cite{HelloVinet,PhysRevA.44.7022}. The residual absorption induces heating of the mirror, which deforms and, at first order, causes a change in the radii of curvature of the mirrors. It thus induces a change in the round-trip ABCD matrix and a modification of the beam size at the interaction point, but also on the mirrors. When different substrates are used \cite{carstens2014megawatt} for input and end mirrors, the apparent symmetry of the cavity is also broken, which can induce a displacement of the cavity waist. It might thus affect the overlap function $\Upsilon$ of the beams at the interaction point. Furthermore, as the OEC mode is changed, the ability to efficiently couple the seed laser to the cavity might be affected \cite{carstens2014megawatt,lu2024stable}. Moreover, mode degeneracies occur, and the related resonant high-order mode must be damped \cite{Amoudry:20}. Finally, in some specific situations, where the spot size on the mirrors is small, the ability to couple the seed laser to the mode of the OEC might be seriously affected by additional thermal lensing in the input mirror \cite{700kW}. In the following, we assume that mode degeneracies are experimentally treated by optimizing the position of D-shaped mirrors specifically inserted into the optical cavity for this purpose. Their position is optimized to preserve the gain for the fundamental mode, while inducing losses to higher-order modes~\cite{10.1063/5.0222951}. By optimizing this position, we checked with the optimized geometry that only Hermite-Gauss modes up to $(3,3)$ are found to have degenerate frequency with the fundamental mode at about 160~kW at 515~nm, similar to what was done in Ref~\cite{PhysRevAccelBeams.25.101601}. Operation at significantly different average power allows for avoiding these degeneracies. Since we intend to have a relatively focused laser beam at the interaction point, the spot size will be relatively large on the mirrors, and we expect the dominant effect to be a cavity mode change rather than thermal lensing. To minimize the influence of thermal loading under these simplifications, one estimates the coupling loss in the presence of thermal effects and intends to minimize it.

The coupling $C_T$ between the electric fields of the seed laser $E_{\text{l}}$ and that circulating into the OEC, $E_{\text{c}}$, reads: 
\begin{equation}
C_T=\frac{\mid\int_{-\infty}^{+\infty}\int_{-\infty}^{+\infty}E_{\text{l}}(x, y)E_{\text{c}}^{*}(x, y)\text{d}x\text{d}y\mid^{2}}{\int_{-\infty}^{+\infty}\int_{-\infty}^{+\infty} \mid E_{\text{l}}(x, y)\mid^{2}\text{d}x\text{d}y\int_{-\infty}^{+\infty}\int_{-\infty}^{+\infty}\mid E_{\text{c}}(x, y) \mid^{2}\text{d}x\text{d}y};
\end{equation} 
where we assumed that clipping losses due to the mirror aperture are negligible, and the contribution from the longitudinal part of the field $(z,t)$ to the coupling factorizes. Assuming a stationary laser pulse envelope, it reduces to a phase noise contribution that is expressed elsewhere and essentially constrains the choice of the laser seeder~\cite{PhysRevAccelBeams.25.101601}. This contribution does not affect the design of the OEC, one thus concentrates on the contribution from the transverse components of the field $C_T$. We assume that the circulating field in the OEC is a fundamental $(0,0)$ Hermite-Gauss mode. For simplicity, to study the influence of thermal loading, we assume that the seed laser also exhibits a fundamental Hermite-Gauss mode. The coupling loss related to higher order mode contributions does not affect the design of the OEC either. These scaling factors only affect the ratio of intracavity and seed laser average powers. As a consequence, the transverse coupling factorizes as the product of two terms $C_T=C_xC_y$, where
\begin{eqnarray}
C_{x/y} & = & \frac{2kw_{0, l, x/y}w_{0, c, x/y}}{\sqrt{k^{2}(w_{0, c, x/y}^{2}+w_{0, l, x/y}^{2})^{2}+4(z_{0, c, x/y}-z_{0, l, x/y})^{2}}}\\
& = & \frac{2w_{1,l,x/y}w_{1,c,x/y}}{\sqrt{(w_{1,l,x/y}^2+w_{1,c,x/y}^2)^2+\frac{k^2\left(\rho_{1,l,x/y}-\rho_{1,c,x/y}\right)^2w_{1,l,x/y}^4w_{1,c,x/y}^4}{\rho_{1,l,x/y}^2\rho_{1,c,x/y}^2}}}
\label{eq:coupling}
\end{eqnarray}
with $E_{\text{l/c}}=E_{\text{l/c,x}}E_{\text{l/c,y}}$ and $E_{\text{l/c,x}} \propto \exp\left(-ikx^2/(2q_{\text{l/c,x}}\right)$ and $q_{\text{l/c,x}}=(z-z_{\text{0, l/c,x}})+ikw^{2}_{\text{0,l/c,x}}/2$ (similarly in $y$ direction). The waist size and position are $w_0$ and $z_0$, respectively. It is also expressed in terms of the actual beam radius $w_{1,l/c,x/y}$ and curvature of the wavefront $\rho_{1,l/c,x/y}$ on the internal face of the OEC of the coupling mirror M$_1$, that we set to $z=0$ by convention.

\begin{figure}[htbp]
    \centering
    \includegraphics[width=0.49\textwidth]{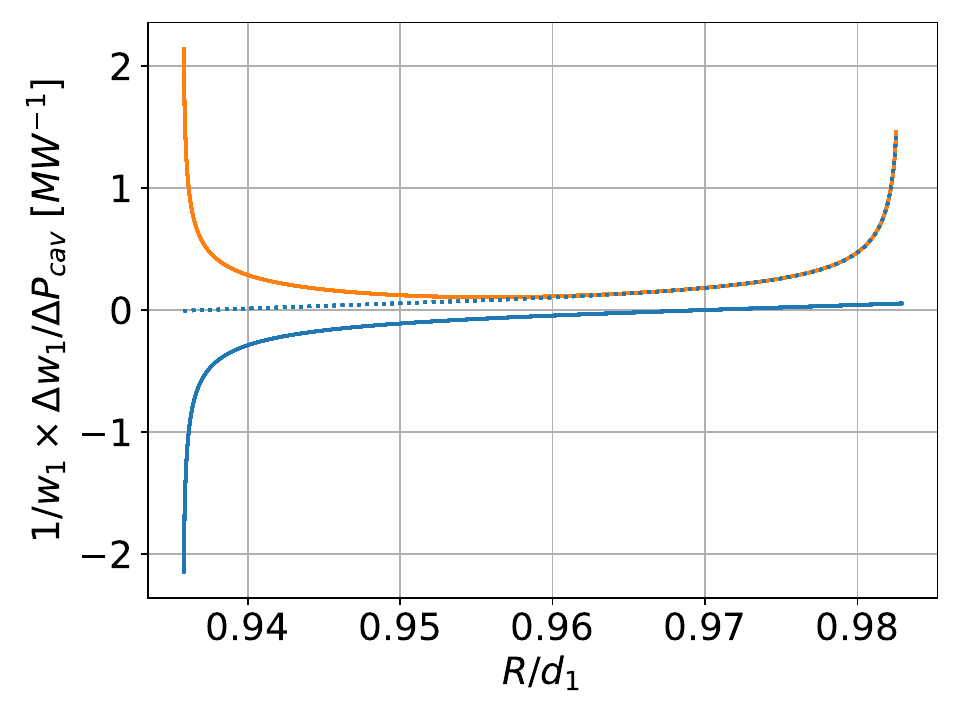}
    \caption{Thermal sensitivity as a function of $R$ with $\lambda_{\text{0}}=\qty{515}{\nm}$. It is defined by the variation of the laser beam size on the input mirror $M_{\text{1}}$ with respect to the average intracavity power $P_{\text{cav}}$. The full (dashed) blue line denotes the relative mode change in the tangential (sagittal) plane. The full orange line denotes the overall thermal sensitivity $\sqrt{(\frac{1}{w_{\text{1, t}}}\frac{\Delta w_{\text{1, t}}}{\Delta P_{\text{cav}}})^{2}+(\frac{1}{w_{\text{1, s}}}\frac{\Delta w_{\text{1, s}}}{\Delta P_{\text{cav}}})^{2}}$. This curve presents a minimum at about $R/d_1\approx 0.955$.}
    \label{fig:thermal_sensitivity}
\end{figure}

\begin{figure}[htbp]
    \includegraphics[width=0.49\textwidth]{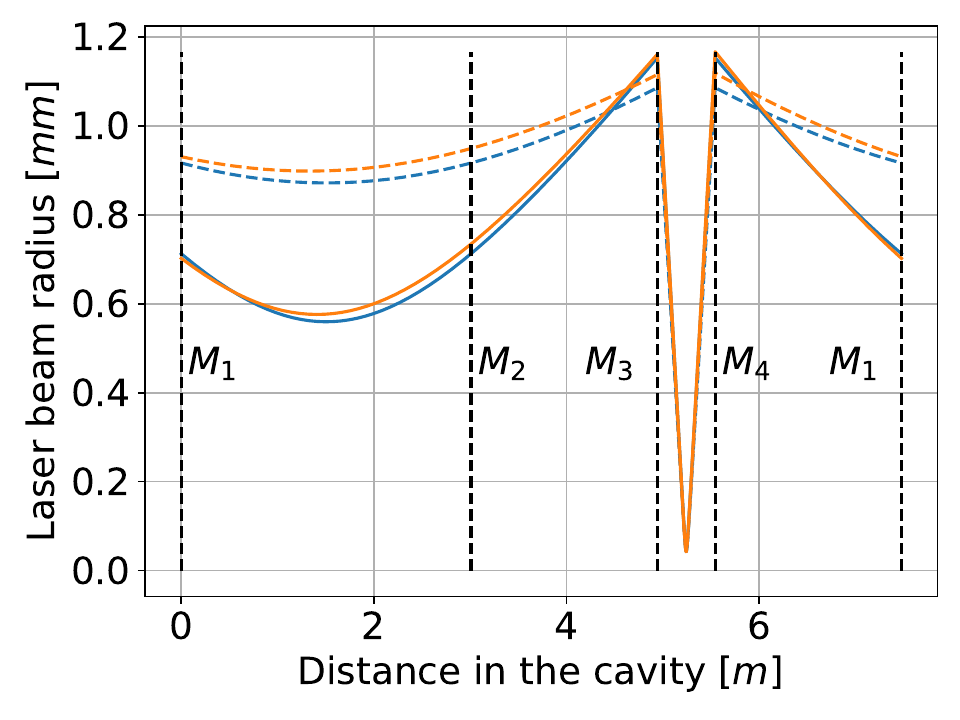}
    \includegraphics[width=0.49\textwidth]{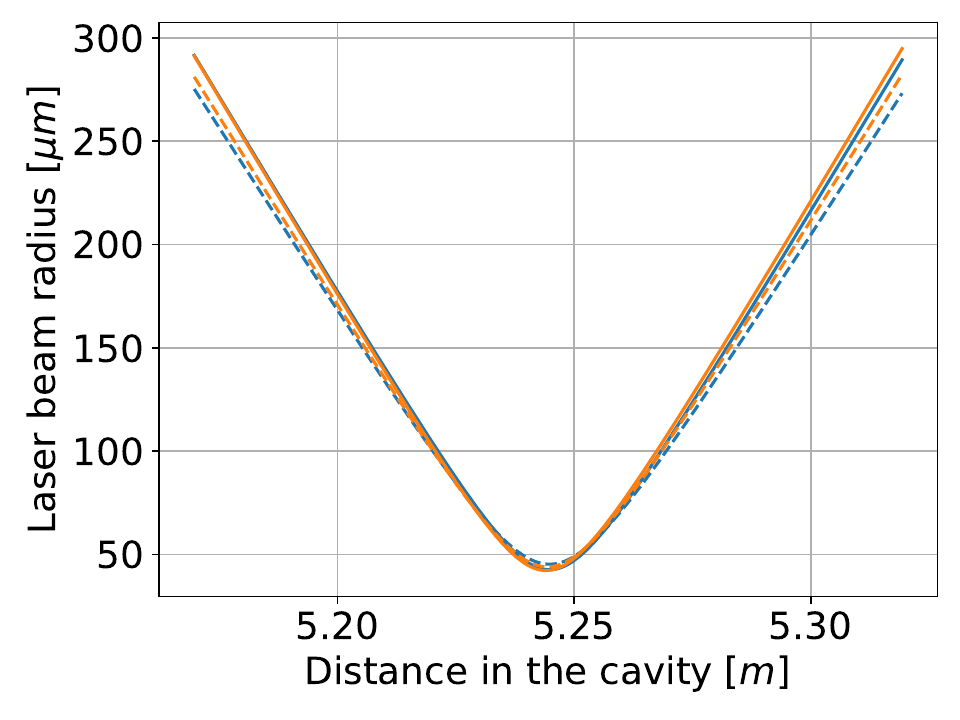}
    \caption{(Left) The OEC fundamental transverse eigenmode laser beam radius as a function of the position in the cavity. The blue lines denote the mode without thermal loading and the orange lines denote the mode computed including thermal loading at $P_{\text{cav}}=\qty{200}{\kW}$. The full lines denote the tangential plane, and the dashed lines denote the sagittal plane. (Right) Zoom in on the region where the electron beam and laser are interacting. The RMS laser intensity is 21.5~$\mu$m in the tangential plane and 22.5~$\mu$m in the sagittal plane.}
    \label{fig:cavity_transverse_fundamental_eigenmode}
\end{figure}

Thermal loading in the OEC is modeled with the Winkler model \cite{PhysRevA.44.7022}. We include the two main contributions that are affecting the operation of this type of optical system \cite{lu2024stable,700kW}. First, thermal expansion of the mirror substrates due to a residual absorption in the coating induce a change of the principal radii of curvature of the mirrors such that $R_{i,k}^{-1}=R_{i,k,0}^{-1}-R_{i,k,\text{th}}^{-1}$ with $R_{i,k,\text{th}}^{-1}=\frac{a_i\alpha_i}{2\pi\kappa_iw_{i,k}^2}P_{\text{cav}}$ where the indices $i={1,2,3,4}$ and $k={x,y}$ denote the mirror index and the axis, respectively. The designed principal radii of curvature are $R_{i,k,0}$, and the parameters $a_i$, $\alpha_i$, and $\kappa_i$ correspond to the coating absorption, the substrate thermal expansion, and thermal conductivity, respectively. We assume for the simulation that $a_1=a_2=a_3=a_4=10^{-6}$~\cite{lma} despite the coating recipes for the coupling mirror and the three other mirrors being different, thus inducing a different $a_1$ for the coupling mirror. Shall this absorption not be reached in practice, it must simply be noted that the results of this paper hold provided that one realizes that the same thermal effects observed with 1~ppm at 200~kW will be observed at 100~kW with a 2~ppm mirror~\cite{layertec}, as only the product of the mirror absorption and the average intracavity power matter here. We assume that the coupling mirror is made of high-grade fused silica and take $\alpha_1=\qty{0.45e-6}{K^{-1}}$, $\kappa_1=\qty{1.38}{W.m^{-1}.K^{-1}}$. The three other mirrors are assumed to be made of ultra-low expansion glass with $\alpha_{2,3,4}=\qty{0.01e-6}{K^{-1}}$ and $\kappa_{2,3,4}=\qty{1.31}{W.m^{-1}.K^{-1}}$. The optical indices are $n_{\text{1}}=1.461$ at 532~nm and $n_{\text{2}}=1.4828$ at 589~nm.~\cite{heraeus, corning} Second, thermal lensing in the coupling can induce coupling loss \cite{700kW}. The effect is modeled by a $P_{\text{cav}}$-dependent focal length $f_i$, for mirror M$_i$, that reads $f_{i,k}^{-1} = \beta_ia_iP_{\text{cav}}/(\pi\kappa_iw_{i,k}^2)$ where the thermo-optic coefficient of the bulk is taken to be $\beta_1 = 8.1\cdot 10^{-6} \text{K}^{-1}$ and $\beta_2 = 10.7\cdot 10^{-6} \text{K}^{-1}$ for Suprasil and ULE, respectively. The lensing in the mirrors M$_{2,3,4}$ is needed to compute beam sizes in transmission of the optical cavity, which is of general interest when comparing simulations with experiments and is thus included, though not affecting the operation of the OEC per se. As the principal radii of curvature in the presence of thermal load depend on the laser beam size on the mirrors, which is computed from the round-trip ABCD matrix of the OEC, a recursive approach is employed. The method described in \cite{10.1063/5.0222951} is implemented. The convergence is generally obtained in a few iterations, and explicitly checked when analyzing the results. As a consequence, the optical cavity mode is expected to change in the presence of thermal loading. Moreover, the apparent symmetry of the optical cavity is also broken since the two planar mirrors suffer different thermal loading as they are made of different materials. The latter effect is, however, small enough and does not affect the calculation of the interaction rate with the electron beam significantly. The numerical values for the geometry parameters are taken to those of \autoref{tab:specs_oec}. The OEC is simulated for various values of the design radius of curvature of the spherical mirrors and for $P_{\text{cav}}=0$ and 200~kW. The change in the two principal radii of curvature of the mirrors $\Delta w_{1,k}=w_{1,k}(P_{\text{cav}}=200~\textrm{kW})-w_{1,k}(P_{\text{cav}}=0~\textrm{kW})$ is computed, which allows to determine $\Delta w_{1,k}/(w_{1,k}(P_{\text{cav}}=0~\textrm{kW}) \Delta P_{\text{cav}})$. We have checked that the beam radius change is indeed linear in this range, by computing $w_{1,k}$ in steps of 50~kW, which is consistent with Winkler's approximation. It is possible to show that if the wavefront curvatures are approximately matched, the coupling change essentially reads $C_T\approx 1-\frac{\delta_{x}^2+\delta_{y}^2}{2}$ where $\delta_{x/y}=\Delta w_{1,x}/w_{1,x}$. This is not exact, but it provides a guide to optimize the design radius of curvature of the mirrors. This thermal sensitivity of the OEC mode is shown in \autoref{fig:thermal_sensitivity}, where $\Delta w_{1,k}/\Delta P_{\text{cav}}$ for $k={x,y}$ is shown along with $\Delta w_{1,k}/\Delta P_{\text{cav}}$. A minimum for the thermal sensitivity is found around $R/(2d_1)\approx 0.955$, thus our choice for the design of PERLE OEC. The cavity fundamental transverse eigenmode propagation is shown in \autoref{fig:cavity_transverse_fundamental_eigenmode} for the cavity in the absence and presence of thermal loading. The position of the mirrors is identified. A zoom around the interaction region is also provided. As expected, we can check that the mode is only slightly changed by the thermal loading. A very small shift of the waist position is observed, which does not affect the interaction luminosity. The slight asymmetry between M$_1$ and M$_2$ is also visible.

\subsection{Influence of thermal lensing on the mode matching}

In order to estimate the sensitivity of the OEC to thermal loading, further assumptions are needed. Indeed, the ability to couple laser power to the optical cavity will highly depend on the design of an appropriate telescope. An intense experimental effort can be invested in this. However, it is likely needed to adjust the telescope while ramping up the optical power inside the OEC \cite{700kW}. There is unfortunately no general answer to this question, as the answer highly depends on the input laser beam and the complexity involved in a specific telescope design. As a consequence, we aim to design a \emph{simple} system that will be relatively insensitive to this effect. In order to do so, we have to specify further the optical telescope and laser beam that will be used to feed laser power to the OEC. We will thus assume, for simplicity, at this early design stage that a two-spherical-lens telescope is employed, similarly to that used in previous work \cite{700kW}. In particular, we do not aim at matching the ellipticity of the OEC mode. The chosen focal lengths are of $f_1=+150$~mm and $f_2=-100$~mm. The assumed waist of the laser is located $d_w=190$~mm right after the amplifier output and exhibits a waist of $w_a=165~\mu$m. It corresponds to the systems used in some previous experiment~\cite{700kW}. The distance in between the amplifier output and the coupling mirror of the OEC is kept fixed at $d_{AM_1}=2.7$~m. Those parameters are summarized in \autoref{tab:specs_laser}. Even though these parameters will have to be updated when the system is bought, the method is informative on the thermal sensitivity of the system. One adjusts the distance $d_{12}$ between the two lenses and the distance $d_{a1}$ between the amplifier output and the first focusing lens.

\begin{figure}[htbp]
    \centering
        \includegraphics[width=0.49\textwidth]{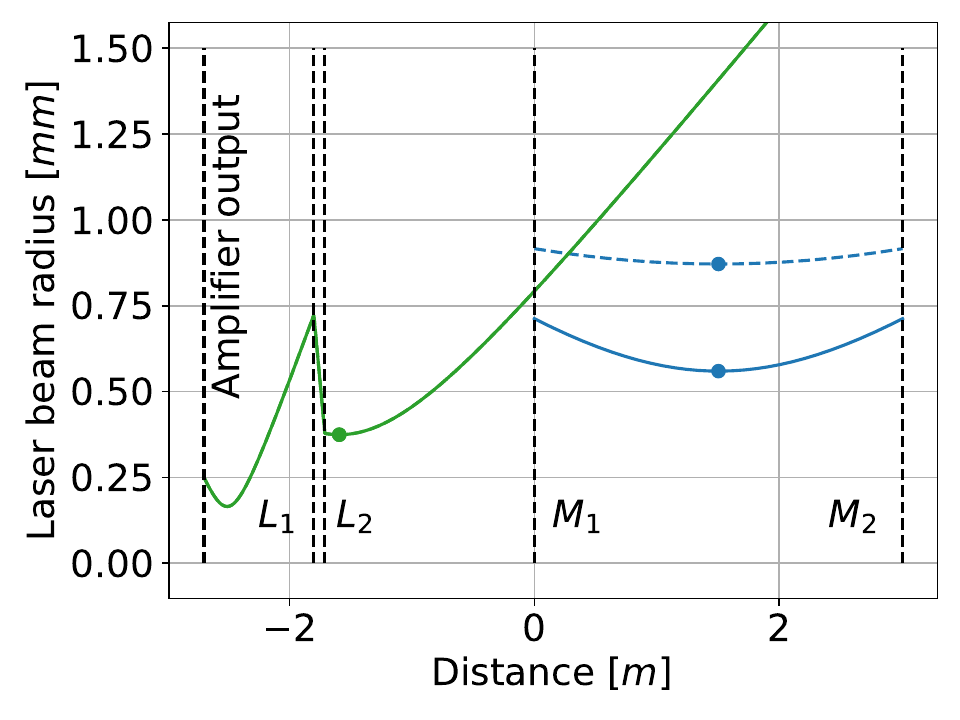}
        \includegraphics[width=0.49\textwidth]{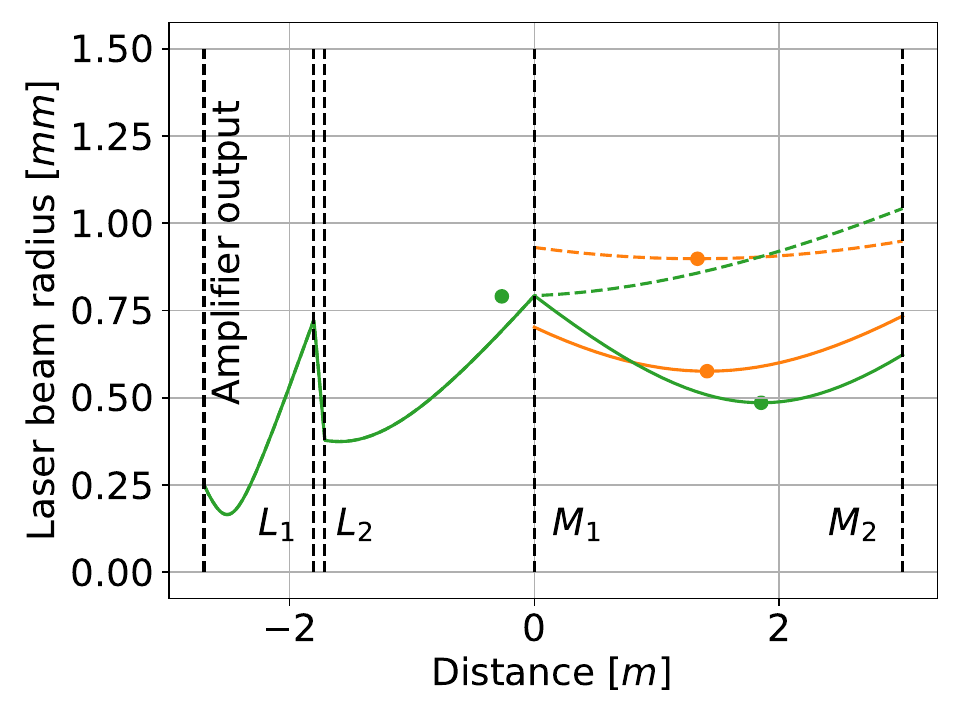}
    \caption{Laser beam radius of the cavity fundamental transverse eigenmode as a function of the position in the optical setup for $\lambda=515nm$ at (left) $P_{\text cav.}=0$~kW  and (right) $P_{\text cav.}=200$~kW. The blue lines (left plot) and orange lines (right plot) denote the cavity eigenmode. The green lines denote the input laser beam radius, including its extrapolation inside the optical cavity, to account for lensing in the input coupler at high average power. The full (dashed) lines represent the beam radius in the tangential (saggital) plane. The dots denote the mode waist and its position, in both directions, for both the cavity fundamental transverse mode and the input laser beam.}
    \label{fig:laser_matching}
\end{figure}

The thermal lensing in the input mirror leads to changes in the incident laser transverse mode, which deteriorates the coupling $C_T$, if not compensated by a different tuning of the matching telescope \cite{700kW}. An example is shown on the \autoref{fig:laser_matching} where the laser beam radius propagation from the amplifier to the OEC is represented. The laser beam delivered by the amplifier is assumed to be a perfect circular fundamental Gaussian beam. As can be seen on the right of the figure, the thermal lensing occurring in M$_1$ generates ellipticity of the input laser beam. One interesting aspect is that in the presence of thermal loading of the cavity, the input laser beam gets a natural ellipticity that tends to better match that of the OEC mode. As a consequence, the thermal loading also has a positive effect on the laser beam coupling for this specific design. This motivates, a posteriori, the idea of not putting efforts into the optimisation of the telescope with cylindrical lenses. This statement is, however, difficult to generalize over the whole range of the stability region as related to the coupling of the seed laser to the intrinsic mode of the cavity. It naturally depends on the effort placed into the telescope design, which is quite minimal here.

\begin{figure}[htbp]
    \centering
        \begin{subfigure}[h]{0.49\textwidth}
            \centering
            \includegraphics[width=\textwidth]{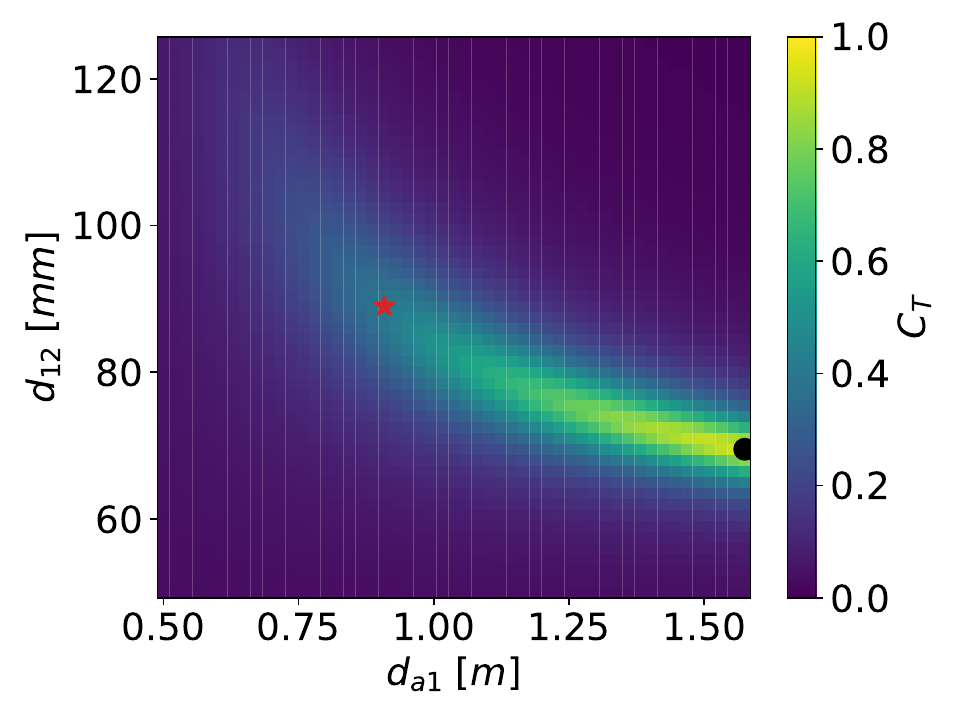}
            \caption{\centering $P_{\text{cav}}=0$~kW.}
            \label{subfig:CT_froid_0515}
        \end{subfigure}
        \hfill
        \begin{subfigure}[h]{0.49\textwidth}
            \centering
            \includegraphics[width=\textwidth]{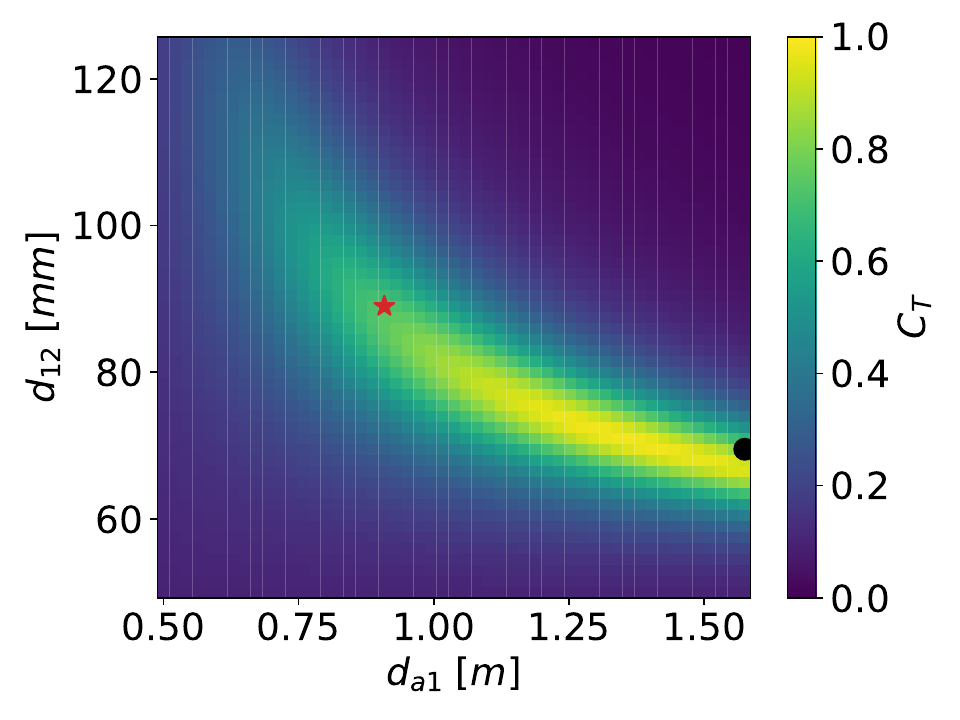}
            \caption{\centering $P_{\text{cav}}=100$~kW.}
            \label{subfig:CT_mid_0515}
        \end{subfigure}
        \hfill
        \begin{subfigure}[h]{0.49\textwidth}
            \centering
            \includegraphics[width=\textwidth]{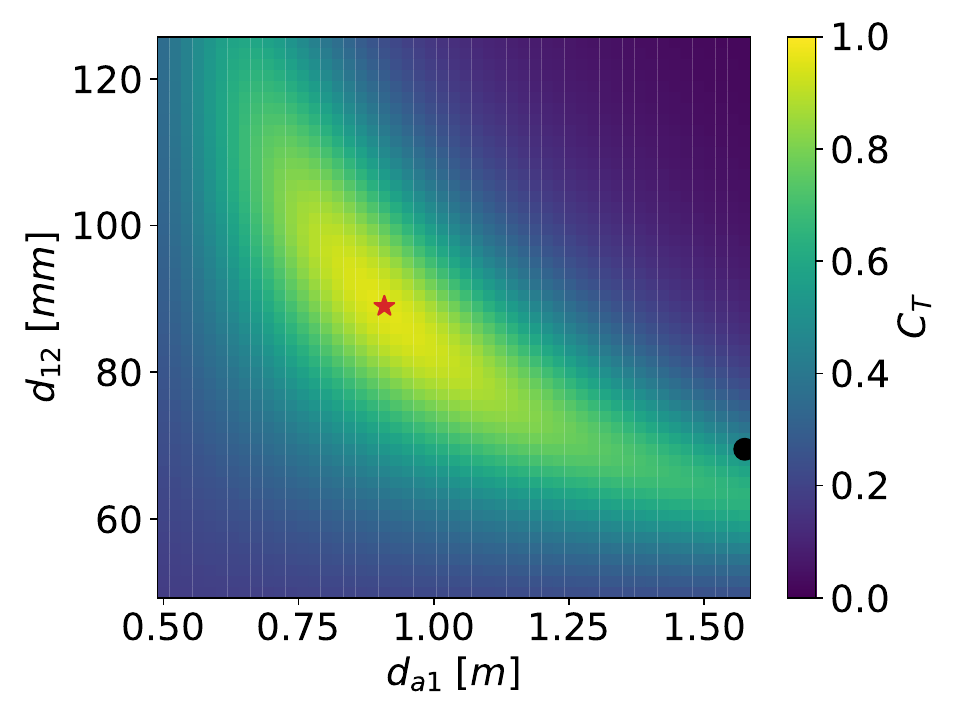}
            \caption{\centering $P_{\text{cav}}=200$~kW.}
            \label{subfig:CT_chaud_0515}
        \end{subfigure}
    \caption{Coupling $C_{\text{T}}$ as a function of the telescope parameters $d_{\text{a1}}$ and $d_{\text{12}}$ for $\lambda=515$~nm for three assumed values of $P_{\text{cav}}$. The black dots (red stars) represent the position of the lenses for the telescope optimized for a cavity at $P_{\text{cav}}=0$~kW ($P_{\text{cav}}=200$~kW).}
    \label{fig:telescope_optimization_0515}
\end{figure}

The sensitivity of the telescope settings is shown in \autoref{fig:telescope_optimization_0515} for three different assumed values of the power in the OEC. A parameter scan is performed assuming that focal lengths are kept fixed, but distances are adjusted. At these moderate average power, it is possible to find a telescope design that allows to match relatively well the laser beam to the OEC mode at 200~kW. It is noticeable that the optimized position of the doublet of lenses is modified significantly by 80~cm, and the distance between the lenses by 2~cm. 

\begin{table}[htbp]
    \centering
    \caption{Laser parameters used during the design. $C_{\text{total}}$ encompasses the transverse coupling, the longitudinal coupling and the polarization coupling between the seed laser and the OEC.} \label{tab:specs_laser}
    \smallskip
    \resizebox{\textwidth}{!}{
    \begin{tabular}{c|c|c|c|c|c|c|c|c|c}
        \hline $\lambda$ [\unit{\nm}] & $U$ [\unit{\mJ}] & $\mathcal{F}$ & G & $f_{rep}$ [\unit{\MHz}] & $\sigma_{\text{z}}$ [\unit{\ps}] & $d_{w}$ [\unit{\mm}] & $w_{a}$ [\unit{\um}] & $P_{\text{las}}$ [\unit{\W}] & $C_{\text{total}}$ \\
        \qty{515}{} or \qty{1030}{} & \qty{5}{} or \qty{25}{} & $\simeq$4800 & $\simeq$5700 & \qty{40.077}{} & \qty{5}{} & \qty{190}{} & \qty{165}{} & \qty{50}{} & 0.7 \\
        \hline
    \end{tabular}}
\end{table}

\section{Performance of the designed OEC}\label{sec:perf}
With the proposed design of the OEC for PERLE, the performance is studied regarding several aspects described below. 

\subsection{Rate of scattered photons as a function of intracavity power}
\begin{figure}[htbp]
    \centering
    \includegraphics[width=0.49\linewidth]{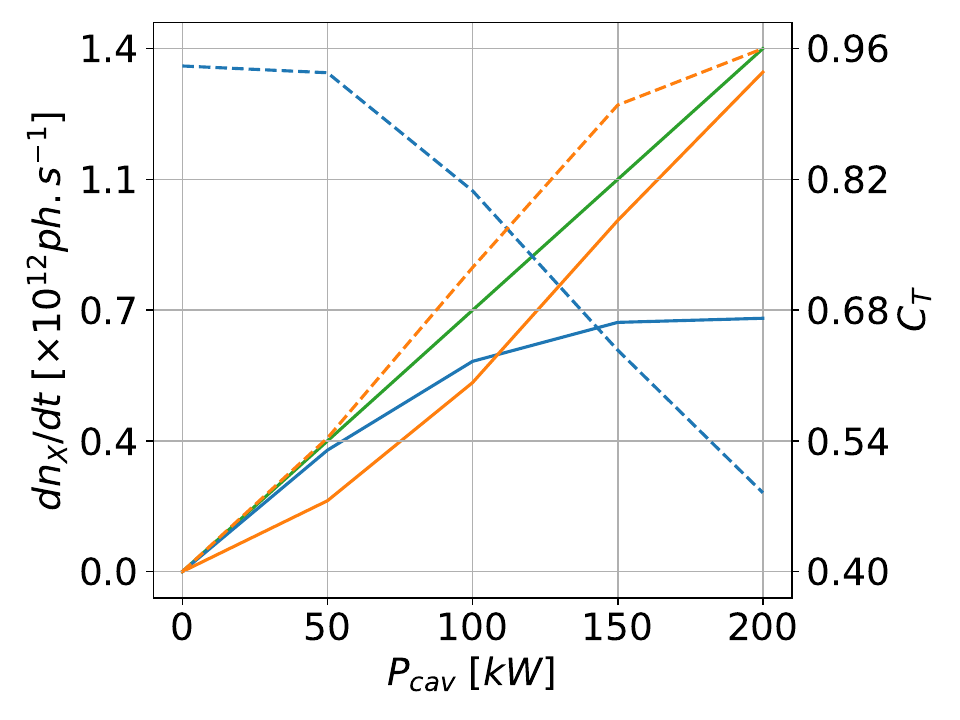}
    \caption{Coupling $C_{\text{T}}$ and photon rate $dn_{\text{X}}/dt$ for $\lambda=515$~nm. The full lines denote the photon rate (left vertical scale), and the dashed lines denote the coupling (right vertical scale). The green line corresponds to the nominal photon rate assuming $C_{\text{T}}=1$. The blue lines correspond to a telescope optimized assuming no thermal loading, and the orange lines correspond to a telescope optimized for $P_{\text{cav}}=200$~kW, including thermal loading.}
    \label{fig:flux515}
\end{figure}

As the design is meant to achieve a stable performance in the presence of thermal loading of the mirror, this aspect is first validated by showing on \autoref{fig:flux515} the rate of scattered photons as a function of the intracavity power. The evolution of the transverse coupling $C_T$ is also provided for two choices of telescope designs, one optimized for low power operation and a second one optimized at the nominal intracavity power. It shows that the nominal rate of photons is not spoiled by expected thermal loading of the OEC. Note that other possible defects, limiting the laser coupling as longitudinal coupling and possible departure of the the input laser beam from the fundamental Gaussian mode assumed here might affect the expected rate of produced photons. Those aspect are a matter of experimental developments are their optimization depend on the experimental realization, kept for the next stage of the project.

\subsection{Performance in infrared}
\begin{figure}[htbp]
    \centering
    \includegraphics[width=0.49\linewidth]{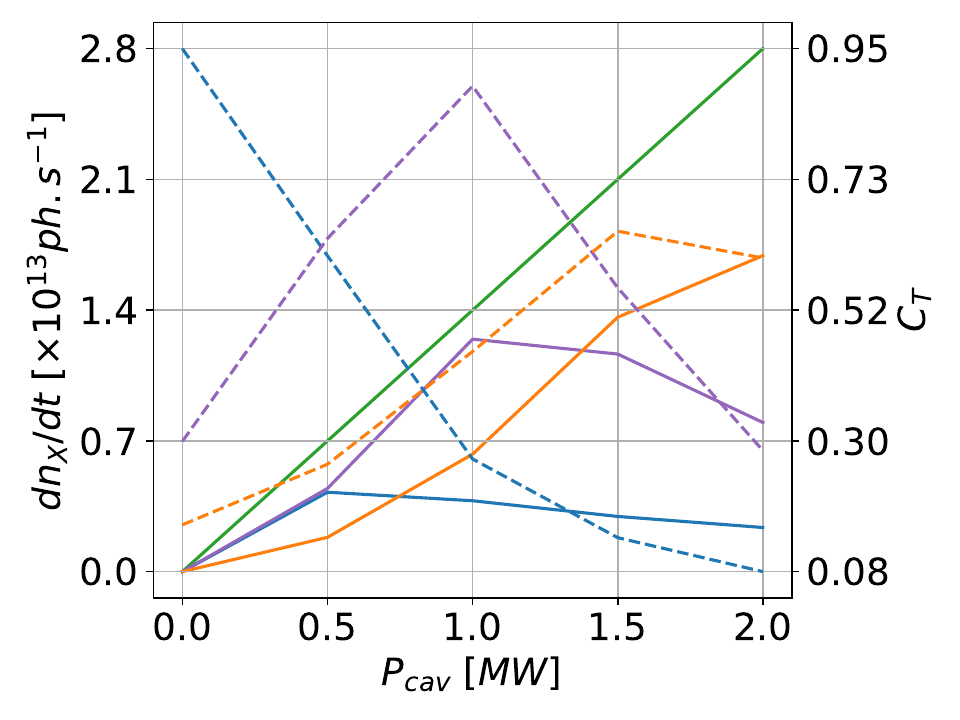}
    \caption{Coupling $C_{\text{T}}$ and photon rate $dn_{\text{X}}/dt$ for $\lambda=1030$~nm. The full lines denote the photon rate (left vertical scale), and the dashed lines denote the coupling (right vertical scale). The green line corresponds to the nominal photon rate with $C_{\text{T}}=1$. The blue lines correspond to a telescope optimized without thermal loading, the purple lines correspond to a telescope optimized for $P_{\text{cav}}=1$~MW, and the orange lines correspond to a telescope optimized for $P_{\text{cav}}=2$~MW.}
    \label{fig:flux1um}
\end{figure}

\begin{figure}[htbp]
    \centering
        \begin{subfigure}[h]{0.49\textwidth}
            \centering
            \includegraphics[width=\textwidth]{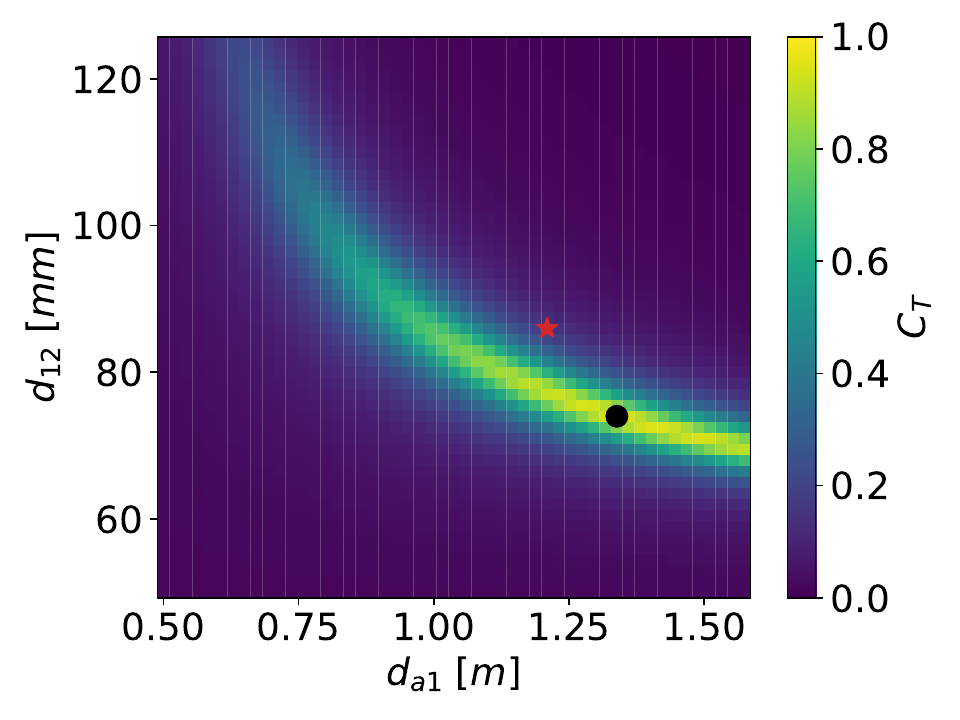}
            \caption{\centering $P_{\text{cav}}=0$~kW.}
            \label{subfig:CT_froid_1030}
        \end{subfigure}
        \hfill
        \begin{subfigure}[h]{0.49\textwidth}
            \centering
            \includegraphics[width=\textwidth]{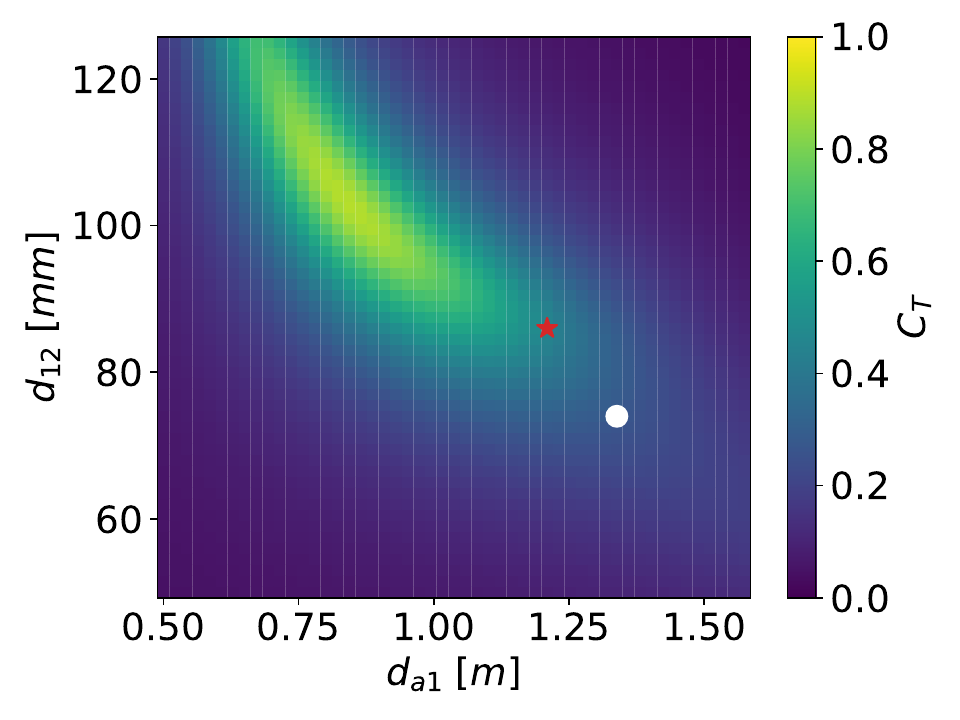}
            \caption{\centering $P_{\text{cav}}=1.0$~MW.}
            \label{subfig:CT_mid_1030}
        \end{subfigure}
        \hfill
        \begin{subfigure}[h]{0.49\textwidth}
            \centering
            \includegraphics[width=\textwidth]{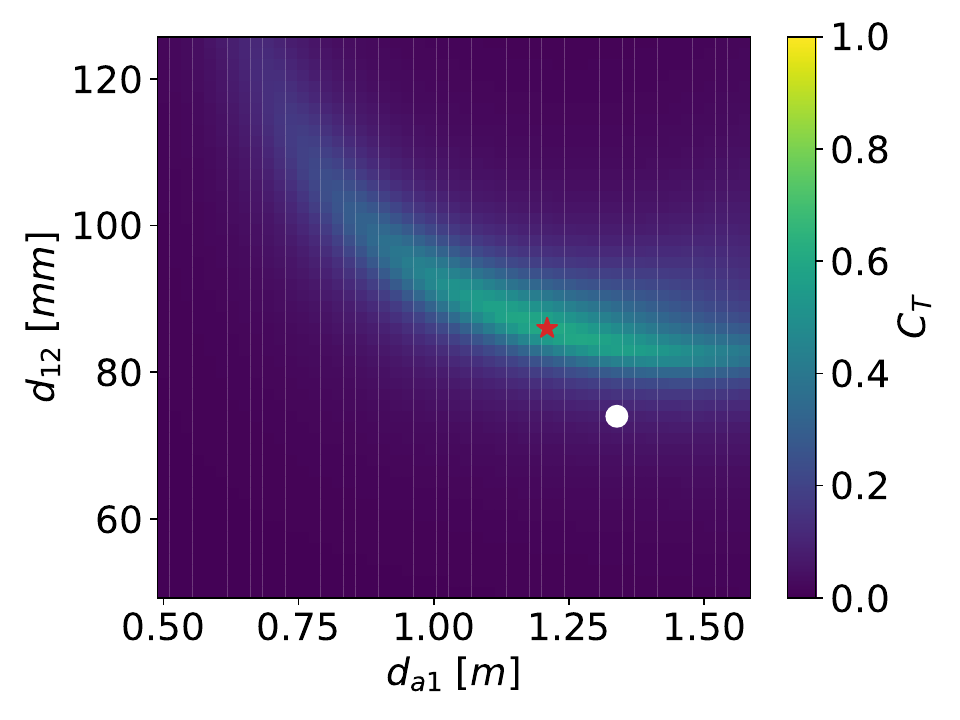}
            \caption{\centering $P_{\text{cav}}=2.0$~MW.}
            \label{subfig:CT_chaud_1030}
        \end{subfigure}
    \caption{Coupling $C_{\text{T}}$ as a function of the telescope parameters $d_{\text{a1}}$ and $d_{\text{12}}$ for $\lambda=1030$~nm. The black and white dots denote the telescope optimized for a cavity without thermal deformations ($P_{\text{cav}}=0$~kW) and the red star denotes the telescope optimized for a cavity with thermal effects ($P_{\text{cav}}=2$~MW).}
    \label{fig:telescope_optimization_1030_1}
\end{figure}

As many existing projects only refer to the use of an OEC operated in infrared, it is interesting to study the performance of this OEC in infrared in the presence of thermal loading. The corresponding curves are shown in \autoref{fig:flux1um}. We deliberately scan the performance well above 1~MW in order to investigate the potential of such a system beyond the near-reach target of 1~MW. The difficulty in matching the laser beam to the OEC mode at 2~MW is striking, and only reflects the simplest choice that could be made for the telescope, a simple double of spherical lenses. This can be seen in the \autoref{fig:telescope_optimization_1030_1}. A quick investigation showed that it was not possible to reach beyond $C_T>0.6$ with such a doublet, even changing the focal lengths. As the influence of thermal lensing becomes stronger at higher thermal loading, it is necessary to investigate, for instance, a quadruplet of cylindrical lenses, which is beyond the scope of this present work. It presents, however, no major difficulty. An elegant alternative solution would consist of replacing the high-grade fused silica with sapphire of excellent surface quality, which remains an open experimental question.

\subsection{Sensitivity to laser pulse duration and crossing angle}
\begin{figure}[htbp]
    \centering
    \includegraphics[width=0.49\textwidth]{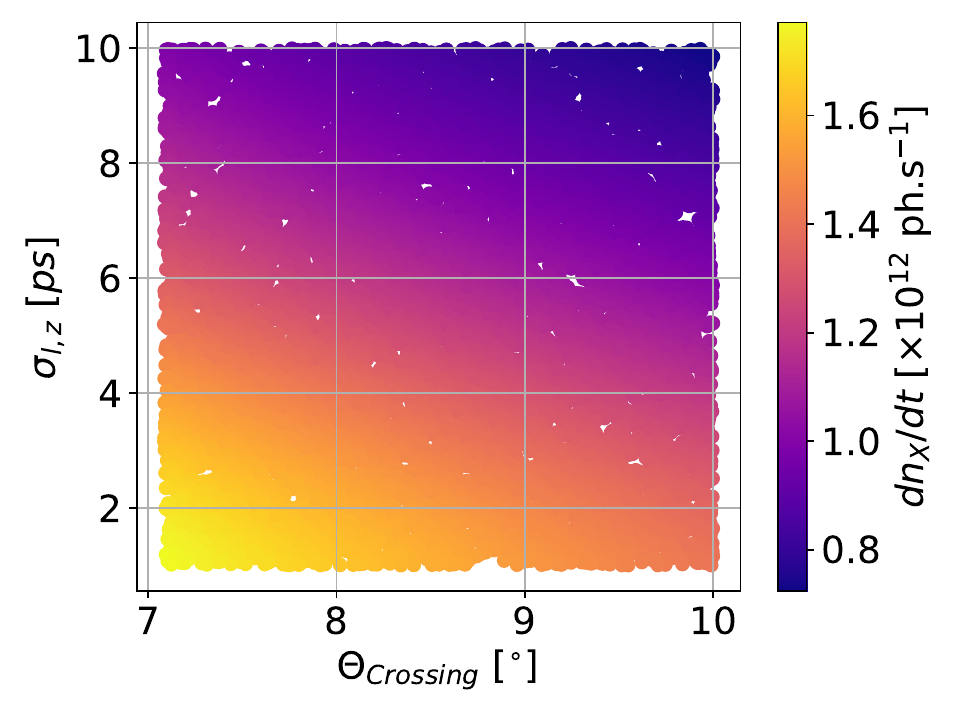}
    \caption{The rate of produced photons as a function of the RMS laser beam pulse duration and the crossing angle between the two beams.}
    \label{fig:optim3}
\end{figure}

The design was made for a given choice of the laser pulse duration. The crossing angle between the laser and electron beams has been fixed to its minimal acceptable value due to integration constraints. However, constraints on this angle might evolve while the project progresses towards implementation. As a consequence, the influence of the crossing angle on the rate of produced photons is studied. Due to the Piwinski contribution, $\sigma_z^2\theta_C^2/\sigma_x^2$ to the luminosity, the impact of the choice of the laser pulse duration is jointly studied, as seen in \autoref{fig:optim3}. It shows an interest in slightly reducing the laser beam pulse duration to a picosecond RMS and keeping as small a crossing angle as possible. Operating with a 10-degree crossing angle would correspond to a 20\% reduction of the scattered photon rate.

\subsection{Alignment sensitivity}
Following the procedure developed in Ref.~\cite{Zomer:09}, we have checked the stability of the optical axis in the OEC. Looking at all possible combinations of $\pm 100~\mu~\!m$ displacements on the four mirrors combined with all possible combinations of $\pm 1$~mrad tilts of the mirrors, the optical axis is found a factor two to four less stable than the ThomX optical cavity \cite{jacquet2024first} due to the much smaller distance between the spherical mirrors. The sensitivity of the cavity to polarization states at high finesse under large incidence angles however remains to be checked in details. Both of these aspects will require careful studies during the implementation of the experimental demonstration.

\subsection{X-ray energy, bandwidth, and brilliance}
\begin{figure}[htbp]
    \centering
    \includegraphics[width=0.49\textwidth]{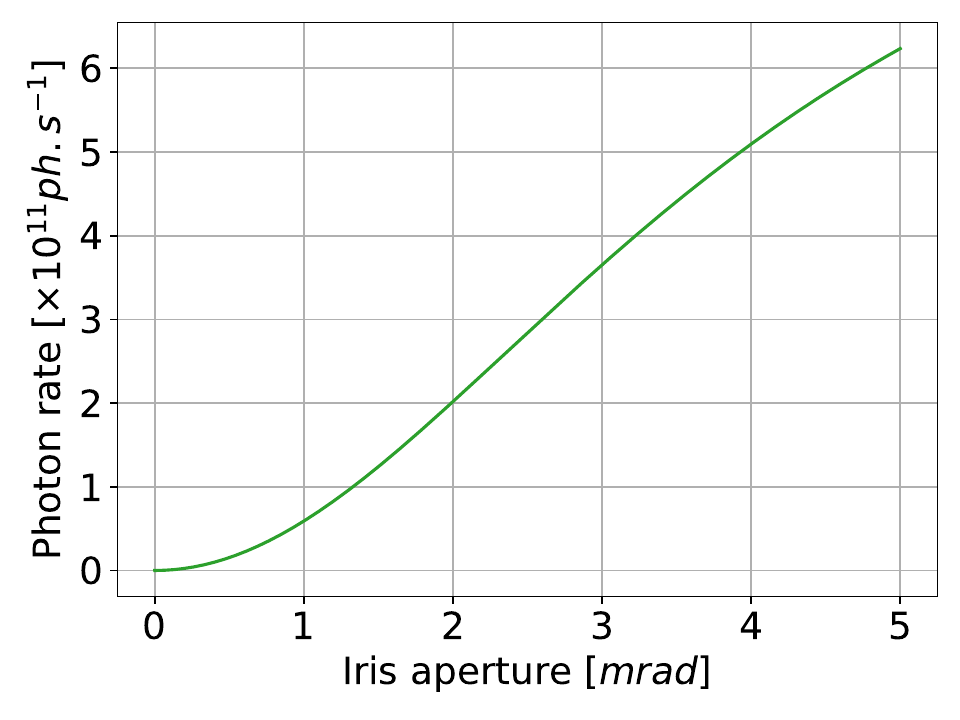}
    \includegraphics[width=0.49\textwidth]{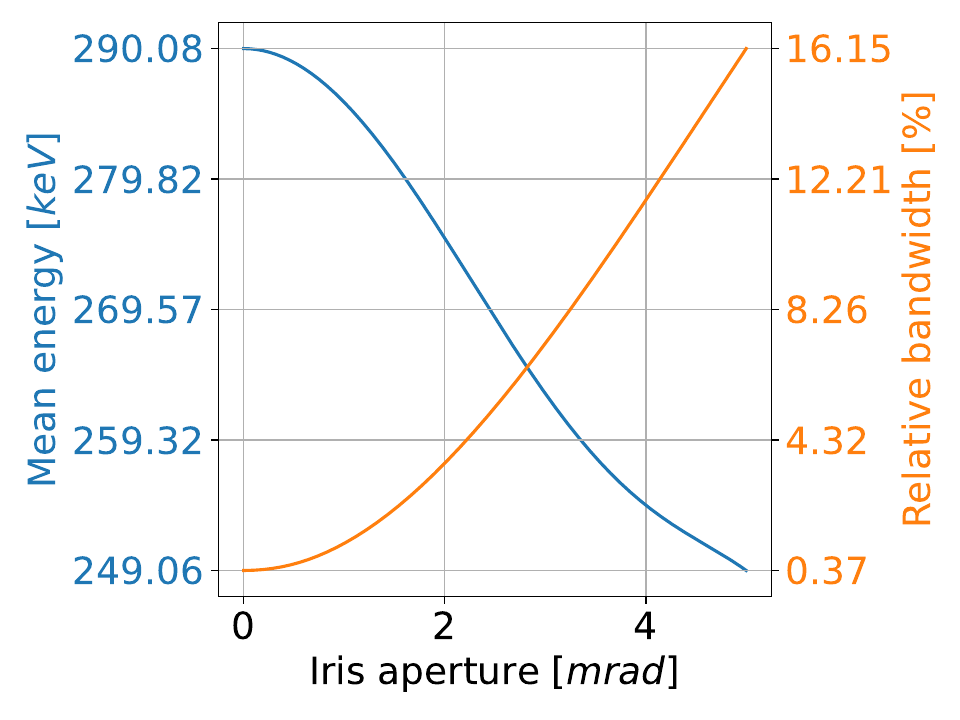}
    \caption{ (Left) X-ray photon rate and (right) mean energy and relative bandwidth for a 515~nm laser as a function of the angular iris aperture. Non-linear corrections are taken into account as described in the text.}
    \label{fig:perf}
\end{figure}

% located 15~m away from the Compton interaction point
One finally estimates, for various angular apertures of a collimating iris, the effective X-ray flux, mean energy, and bandwidth in this aperture using formulae provided in \cite{martens2021towards} and references therein. Simplified expressions for small apertures are derived here to capture the main effects, though the detailed formulae are implemented in the numerical estimation. The relative correction related to the simplified expression is inducing a relative change of quoted values by less than 1\% for this cavity design and the PERLE parameters listed in \autoref{tab:specs_perle}. The rate of scattered X-ray $n_a$, the mean energy of scattered photons $\mu_a$ and the relative bandwidth $\sigma_{a}/\mu_a$ in the aperture approximately read
\begin{eqnarray}
    \label{eq:perf}
	n_a & \approx & \frac{3}{2}\frac{\Psi_0^2}{1+2\Psi_0^2}\frac{dn_{\text{X}}}{dt}\\
    \mu_a & \approx & \frac{E_eX_0}{2(1+X_0)}\left(2-\Psi_0^2\right)\\
    \frac{\sigma_a}{\mu_a} & \approx & \sqrt{\left(\frac{\Psi_0^2}{\sqrt{12}(1+X_0+\Psi_0^2/2)}+\frac{2}{1+X_0}\frac{\epsilon_\text{n}^2}{\sigma_{x,e}^2}\right)^2+\left(\frac{2+X_0}{1+X_0+\Psi_0^2}\frac{\Delta E_e}{E_e}\right)^2}; \text{ where}\\
    X_0 & = &  2(1+\beta\cos\theta_{Crossing})\frac{h\nu_0E_e}{m_e^2c^4 } \approx 3.3\cdot 10^{-3}; \text{ and }\\
    \Psi_0 & = & \frac{\gamma\theta_a}{\sqrt{1+X_0}}.\\
\end{eqnarray}
It is noticeable that the PERLE X-ray source performance is fully limited by the performance of the provided electron beam parameters.
The brightness of the source is also computed  \cite{JACQUET20141} as
\begin{equation}
    \label{eq:brightness}
	\mathcal{B}\approx0.0015\frac{dn_{\text{X}}}{dt}\frac{\sqrt{\sigma_{y,e}^2+\sigma_{y,l}^2}}{\sigma_{y,l}}\frac{\sqrt{\sigma_{x,e}^2+\tilde{\sigma}_{x,l}^2}}{\tilde{\sigma}_{x,l}}\frac{\gamma^2}{(2\pi)^2\epsilon_\text{n}^2} \approx 0.7 \cdot 10^{12}~\text{ph.s$^{-1}$mrad$^{-2}$mm$^{-2}$(0.1\%bw)$^{-1}$},
\end{equation}
where 
\begin{equation}
	\tilde{\sigma}_{x,l}^2 = \sigma_{x,l}^2 + \tan^2{\left(\frac{\theta_{ \textrm{\small Crossing}}}{2}\right)}\left(\sigma_{z,e}^2 +\sigma_{z,l}^2 \right),
\end{equation}
for the 200~kW OEC operated at 515~nm. Note that this expression is slightly improved from that of Ref.~\cite{JACQUET20141} to account for the relatively large crossing angle and that the effective X-ray spot at the interaction point is different in horizontal and vertical directions. The large crossing angle reduces the brightness of the source by a factor of about 3.5. Should more space be available, the mechanical design of the source must be adapted to minimize the crossing angle to improve the source brightness significantly and the production rate of X-ray photons. The estimated performance is shown in \autoref{fig:perf}.

\section{Conclusion}
A conceptual design for an optical enhancement cavity to be operated at high average power at 515~nm has been drawn in the context of the PERLE project. A 4-mirror bow-tie optical cavity is chosen for its stability and ease of operation in an accelerator environment where cavity length and spot size at interaction might be tuned. It accounts for the severe mechanical integration constraints in the ERL environment at the current stage of the project. Thermal loading of the mirror has been accounted for in a three-step process. We show, in this paper, that performance can be preserved at high power by (i) optimizing the geometry ignoring thermal effects; (ii) choosing the radius of curvature of the spherical mirrors by minimizing fundamental mode deformation due to thermal expansion; (iii) recovering coupling by proper design of a simplified matching telescope at high-average power. For the proposed design, we found that the thermal lensing in the input mirror allows for compensation of the mismatch of the circular laser beam to the naturally elliptical mode of the OEC, as lensing strength is different in the two principal axes of the OEC. A total rate of $1.4\cdot 10^{12}$ photons per second, or $2.0\cdot 10^{11}$ within a 2~mrad aperture, is estimated with a brightness of $0.7 \cdot 10^{12}\text{ph.s$^{-1}$mrad$^{-2}$mm$^{-2}$/(0.1\%bw)}$, assuming an average power of 200~kW in the optical cavity at 40~MHz. This simple design methodology is also of more general interest for other projects where thermal loading might be an issue, as a steady state microbunching-based radiation source \cite{10.1063/5.0222951,deng2021experimental}.

\section*{Data availability statements}
The code used to generate data that supports the findings of this article is available under CC-BY-NC 4.0 licence~\cite{gitlabCode}. %For the purpose of open access, the author has applied a CC-BY-NC 4.0 public copyright licence to any Author Accepted Manuscript (AAM) version arising from this submission.

\bibliographystyle{JHEP}
\bibliography{biblio.bib}

\providecommand{\href}[2]{#2}\begingroup\raggedright\begin{thebibliography}{10}

\bibitem{abbott2016gw150914}
B.P.~Abbott, R.~Abbott, T.~Abbott, M.~Abernathy, F.~Acernese, K.~Ackley et~al.,
  \emph{{GW150914: The Advanced LIGO detectors in the era of first
  discoveries}}, {\emph{Physical review letters} {\bfseries 116} (2016)
  131103}.

\bibitem{acernese2014advanced}
F.~Acernese, M.~Agathos, K.~Agatsuma, D.~Aisa, N.~Allemandou, A.~Allocca
  et~al., \emph{{Advanced Virgo: a second-generation interferometric
  gravitational wave detector}}, {\emph{Classical and Quantum Gravity}
  {\bfseries 32} (2014) 024001}.

\bibitem{akutsu_kagra_2019}
T.~Akutsu, M.~Ando, K.~Arai, Y.~Arai, S.~Araki, A.~Araya et~al., \emph{{KAGRA:
  2.5 generation interferometric gravitational wave detector}},
  \href{https://doi.org/10.1038/s41550-018-0658-y}{\emph{Nature Astronomy}
  {\bfseries 3} (2019) 35}.

\bibitem{Willke_2006}
B.~Willke, P.~Ajith, B.~Allen, P.~Aufmuth, C.~Aulbert, S.~Babak et~al.,
  \emph{{The GEO-HF project}},
  \href{https://doi.org/10.1088/0264-9381/23/8/S26}{\emph{Classical and Quantum
  Gravity} {\bfseries 23} (2006) S207}.

\bibitem{B_hre_2013}
R.~Bähre, B.~Döbrich, J.~Dreyling-Eschweiler, S.~Ghazaryan, R.~Hodajerdi,
  D.~Horns et~al., \emph{{Any light particle search II — Technical Design
  Report}}, \href{https://doi.org/10.1088/1748-0221/8/09/t09001}{\emph{Journal
  of Instrumentation} {\bfseries 8} (2013) T09001–T09001}.

\bibitem{PhysRevLett.132.191002}
J.~Heinze, A.~Gill, A.~Dmitriev, J.c.v.~Smetana, T.~Yan, V.~Boyer et~al.,
  \emph{{First Results of the Laser-Interferometric Detector for Axions
  (LIDA)}}, \href{https://doi.org/10.1103/PhysRevLett.132.191002}{\emph{Phys.
  Rev. Lett.} {\bfseries 132} (2024) 191002}.

\bibitem{krasny2015gammafactoryproposalcern}
M.W.~Krasny, \emph{{The Gamma Factory proposal for CERN}},  2015.

\bibitem{PhysRevAccelBeams.25.101601}
A.~Martens, K.~Cassou, R.~Chiche, K.~Dupraz, D.~Nutarelli, Y.~Peinaud et~al.,
  \emph{{Design of the optical system for the Gamma Factory proof of principle
  experiment at the CERN Super Proton Synchrotron}},
  \href{https://doi.org/10.1103/PhysRevAccelBeams.25.101601}{\emph{Phys. Rev.
  Accel. Beams} {\bfseries 25} (2022) 101601}.

\bibitem{simonin2016negative}
A.~Simonin, R.~Agnello, S.~Béchu, J.~Bernard, C.~Blondel, J.-P.~Boeuf et~al.,
  \emph{{Negative ion source development for a photoneutralization based
  neutral beam system for future fusion reactors}}, {\emph{New Journal of
  Physics} {\bfseries 18} (2016) 125005}.

\bibitem{sunahara2025laser}
A.~Sunahara, G.~Raj, T.~Cohen, P.M.~Pattison, P.~Rudy, Y.~Ohara et~al.,
  \emph{{Laser-based inertial fusion energy system enabled by optical
  enhancement cavities and a direct-drive configuration reactor}},
  {\emph{Optics Express} {\bfseries 33} (2025) 47104}.

\bibitem{deng2021experimental}
X.~Deng, A.~Chao, J.~Feikes, A.~Hoehl, W.~Huang, R.~Klein et~al.,
  \emph{{Experimental demonstration of the mechanism of steady-state
  microbunching}}, {\emph{Nature} {\bfseries 590} (2021) 576}.

\bibitem{kruschinski_confirming_2024}
A.~Kruschinski, X.~Deng, J.~Feikes, A.~Hoehl, R.~Klein, J.~Li et~al.,
  \emph{{Confirming the theoretical foundation of steady-state microbunching}},
  \href{https://doi.org/10.1038/s42005-024-01657-y}{\emph{Communications
  Physics} {\bfseries 7} (2024) 160}.

\bibitem{eggl2016munich}
E.~Eggl, M.~Dierolf, K.~Achterhold, C.~Jud, B.~Günther, E.~Braig et~al.,
  \emph{{The Munich compact light source: initial performance measures}},
  {\emph{Journal of synchrotron radiation} {\bfseries 23} (2016) 1137}.

\bibitem{DUPRAZ2020100051}
K.~Dupraz, M.~Alkadi, M.~Alves, L.~Amoudry, D.~Auguste, J.-L.~Babigeon et~al.,
  \emph{{The ThomX ICS source}},
  \href{https://doi.org/https://doi.org/10.1016/j.physo.2020.100051}{\emph{Physics
  Open} {\bfseries 5} (2020) 100051}.

\bibitem{RAKHMAN201682}
A.~Rakhman, M.~Hafez, S.~Nanda, F.~Benmokhtar, A.~Camsonne, G.~Cates et~al.,
  \emph{{A high-finesse Fabry–Perot cavity with a frequency-doubled green
  laser for precision Compton polarimetry at Jefferson Lab}},
  \href{https://doi.org/https://doi.org/10.1016/j.nima.2016.03.085}{\emph{Nuclear
  Instruments and Methods in Physics Research Section A: Accelerators,
  Spectrometers, Detectors and Associated Equipment} {\bfseries 822} (2016)
  82}.

\bibitem{carstens2014megawatt}
H.~Carstens, N.~Lilienfein, S.~Holzberger, C.~Jocher, T.~Eidam, J.~Limpert
  et~al., \emph{{Megawatt-scale average-power ultrashort pulses in an
  enhancement cavity}}, {\emph{Optics letters} {\bfseries 39} (2014) 2595}.

\bibitem{lu2024stable}
X.-Y.~Lu, R.~Chiche, K.~Dupraz, F.~Johora, A.~Martens, D.~Nutarelli et~al.,
  \emph{{Stable 500 kW average power of infrared light in a finesse 35 000
  enhancement cavity}}, {\emph{Applied Physics Letters} {\bfseries 124} (2024)
  }.

\bibitem{700kW}
X.-Y.~Lu, R.~Chiche, K.~Dupraz, A.~Martens, D.~Nutarelli, V.~Soskov et~al.,
  \emph{{710 kW stable average-power in a 45,000 finesse two-mirror optical
  cavity}}, \href{https://doi.org/10.1364/OL.543388}{\emph{Optics Letters}
  {\bfseries 49} (2024) }.

\bibitem{10.1117/12.3042333}
N.~Rīgere, M.~Wurzer, B.~G{\"u}nther, C.~Xue and R.~Kienberger,
  \emph{{Advanced green cavity design for generating high-energy
  partially-coherent x-rays}},  in \emph{Laser Resonators, Microresonators, and
  Beam Control XXVII}, V.S.~Ilchenko, A.M.~Armani, J.V.~Sheldakova,
  A.V.~Kudryashov and A.B.~Matsko, eds., vol.~13349, p.~1334903, International
  Society for Optics and Photonics, SPIE, 2025,
  \href{https://doi.org/10.1117/12.3042333}{DOI}.

\bibitem{graves:linac2024-weya004}
W.~Graves et~al., \emph{{Results from CXLS commissioning}},  in \emph{Proc.
  32nd Linear Accelerator Conference (LINAC2024)}, no.~32 in International
  Linear Accelerator Conference, pp.~557--562, JACoW Publishing, Geneva,
  Switzerland, 08, 2024,
  \href{https://doi.org/10.18429/JACoW-LINAC2024-WEYA004}{DOI}.

\bibitem{VanElk:25}
I.J.M.V.~Elk, C.W.~Sweers, D.F.J.~Nijhof, R.G.W.V.~den Berg, T.G.~Lucas,
  X.F.D.~Stragier et~al., \emph{{First x-rays from a compact and tunable
  LINAC-based Compton scattering source}},
  \href{https://doi.org/10.1364/OE.566542}{\emph{Opt. Express} {\bfseries 33}
  (2025) 47498}.

\bibitem{SAMSAM2024168990}
S.~Samsam, L.~Serafini, M.~Ruijter, F.~Prelz, M.~{Rossetti Conti}, A.~Bacci
  et~al., \emph{{Progress in the energy upgrade of the Southern European
  Thomson back-scattering source (STAR)}},
  \href{https://doi.org/https://doi.org/10.1016/j.nima.2023.168990}{\emph{Nuclear
  Instruments and Methods in Physics Research Section A: Accelerators,
  Spectrometers, Detectors and Associated Equipment} {\bfseries 1059} (2024)
  168990}.

\bibitem{AMOUDRY2025170287}
L.~Amoudry, M.~Kravchenko, R.~Berry, N.~Burger, A.~Diego, J.~Edelen et~al.,
  \emph{{Commissioning of a photocathode and interaction laser system at
  RadiaBeam compact inverse Compton light source}},
  \href{https://doi.org/https://doi.org/10.1016/j.nima.2025.170287}{\emph{Nuclear
  Instruments and Methods in Physics Research Section A: Accelerators,
  Spectrometers, Detectors and Associated Equipment} {\bfseries 1075} (2025)
  170287}.

\bibitem{10.3389/fphy.2024.1472759}
C.P.J.~Barty, J.M.~Algots, A.J.~Amador, J.C.R.~Barty, S.M.~Betts,
  M.A.~Castañeda et~al., \emph{{Design, construction, and test of compact,
  distributed-charge, X-band accelerator systems that enable image-guided, VHEE
  FLASH radiotherapy}},
  \href{https://doi.org/10.3389/fphy.2024.1472759}{\emph{Frontiers in Physics}
  {\bfseries Volume 12 - 2024} (2024) }.

\bibitem{Zhao:17}
Z.~Zhao, B.~Sheehy and M.~Minty, \emph{{Generation of 180 W average green power
  from a frequency-doubled picosecond rod fiber amplifier}},
  \href{https://doi.org/10.1364/OE.25.008138}{\emph{Opt. Express} {\bfseries
  25} (2017) 8138}.

\bibitem{Bullington:08}
A.L.~Bullington, B.T.~Lantz, M.M.~Fejer and R.L.~Byer, \emph{{Modal frequency
  degeneracy in thermally loaded optical resonators}},
  \href{https://doi.org/10.1364/AO.47.002840}{\emph{Appl. Opt.} {\bfseries 47}
  (2008) 2840}.

\bibitem{Bonis_2012}
J.~Bonis, R.~Chiche, R.~Cizeron, M.~Cohen, E.~Cormier, P.~Cornebise et~al.,
  \emph{{Non-planar four-mirror optical cavity for high intensity gamma ray
  flux production by pulsed laser beam Compton scattering off GeV-electrons}},
  \href{https://doi.org/10.1088/1748-0221/7/01/P01017}{\emph{Journal of
  Instrumentation} {\bfseries 7} (2012) P01017}.

\bibitem{suzuki1976general}
T.~Suzuki et~al., \emph{{General formulae of luminosity for various types of
  colliding beam machines}},  1976.

\bibitem{Zomer:09}
F.~Zomer, Y.~Fedala, N.~Pavloff, V.~Soskov and A.~Variola, \emph{{Polarization
  induced instabilities in external four-mirror Fabry-Perot cavities}},
  \href{https://doi.org/10.1364/AO.48.006651}{\emph{Appl. Opt.} {\bfseries 48}
  (2009) 6651}.

\bibitem{PhysRevAccelBeams.21.121601}
P.~Favier, L.~Amoudry, K.~Cassou, R.~Chiche, K.~Dupraz, A.~Martens et~al.,
  \emph{{Optimization of a Fabry-Perot cavity operated in burst mode for
  Compton scattering experiments}},
  \href{https://doi.org/10.1103/PhysRevAccelBeams.21.121601}{\emph{Phys. Rev.
  Accel. Beams} {\bfseries 21} (2018) 121601}.

\bibitem{kogelnik1966laser}
H.~Kogelnik and T.~Li, \emph{{Laser beams and resonators}}, {\emph{Applied
  optics} {\bfseries 5} (1966) 1550}.

\bibitem{HelloVinet}
P.~Hello and J.-Y.~Vinet, \emph{{Analytical models of thermal aberrations in
  massive mirrors heated by high power laser beams}},
  \href{https://doi.org/10.1051/jphys:0199000510120126700}{\emph{J. Phys.
  France} {\bfseries 51} (1990) 1267}.

\bibitem{PhysRevA.44.7022}
W.~Winkler, K.~Danzmann, A.~R\"udiger and R.~Schilling, \emph{{Heating by
  optical absorption and the performance of interferometric gravitational-wave
  detectors}}, \href{https://doi.org/10.1103/PhysRevA.44.7022}{\emph{Phys. Rev.
  A} {\bfseries 44} (1991) 7022}.

\bibitem{Amoudry:20}
L.~Amoudry, H.~Wang, K.~Cassou, R.~Chiche, K.~Dupraz, A.~Martens et~al.,
  \emph{{Modal instability suppression in a high-average-power and high-finesse
  Fabry--Perot cavity}},
  \href{https://doi.org/10.1364/AO.59.000116}{\emph{Appl. Opt.} {\bfseries 59}
  (2020) 116}.

\bibitem{10.1063/5.0222951}
X.~Liu, X.-Y.~Lu, Q.-L.~Tian, Z.-L.~Pan, X.-J.~Deng, L.-X.~Yan et~al.,
  \emph{{Prototype optical enhancement cavity for steady-state microbunching}},
  \href{https://doi.org/10.1063/5.0222951}{\emph{Review of Scientific
  Instruments} {\bfseries 95} (2024) 103004}.

\bibitem{lma}
{Private communication with Laboratoire des Matériaux Avancés
  (CNRS/IP2I/LMA)}.

\bibitem{layertec}
{Layertec website}.
  \url{https://www.layertec.de/en/components/mirrors/low-loss/}.

\bibitem{heraeus}
{Heraeus website}.
  \url{https://www.heraeus-covantics.com/media/Media/Documents/Products_and_Solutions/OPT/EN/Data_and_Properties_Optics_fused_silica_EN.pdf}.

\bibitem{corning}
{Corning website}.
  \url{https://www.corning.com/media/worldwide/csm/documents/7972%20ULE%20Product%20Information%20Jan%202016.pdf}.

\bibitem{jacquet2024first}
M.~Jacquet et~al., \emph{{First production of X-rays at the ThomX
  high-intensity Compton source}}, {\emph{The European Physical Journal Plus}
  {\bfseries 139} (2024) 1}.

\bibitem{martens2021towards}
A.~Martens, F.~Zomer, M.~Amer, L.~Amoudry, K.~Cassou, K.~Dupraz et~al.,
  \emph{{Towards ultimate bandwidth photon sources based on Compton
  backscattering: Design constraints due to nonlinear effects}},
  {\emph{Physical Review Accelerators and Beams} {\bfseries 24} (2021) 091601}.

\bibitem{JACQUET20141}
M.~Jacquet, \emph{{High intensity compact Compton X-ray sources: Challenges and
  potential of applications}},
  \href{https://doi.org/https://doi.org/10.1016/j.nimb.2013.10.078}{\emph{Nuclear
  Instruments and Methods in Physics Research Section B: Beam Interactions with
  Materials and Atoms} {\bfseries 331} (2014) 1}.

\bibitem{gitlabCode}
A.~Renaux et~al., ``{Software used for the design of the OEC presented in this
  paper, under CC-BY-NC 4.0 licence}.''
  \url{https://gitlab.in2p3.fr/alice.renaux/HPOEC-model}.

\end{thebibliography}\endgroup

\end{document}